\documentclass[fleqn,usenatbib]{mnras}

\usepackage{newtxtext,newtxmath}

\usepackage[T1]{fontenc}

\DeclareRobustCommand{\VAN}[3]{#2}
\let\VANthebibliography\thebibliography
\def\thebibliography{\DeclareRobustCommand{\VAN}[3]{##3}\VANthebibliography}

\usepackage{graphicx}	
\usepackage{amsmath}	
\usepackage{amsmath}	
\usepackage{adjustbox} 
\usepackage{tikz} 
\usepackage{lscape}  
\usepackage{setspace,caption} 
\usepackage{lineno} 
\usepackage{graphics,graphicx}
\usepackage[utf8]{inputenc}
\usepackage[english]{babel}

\newcommand{\kms}{\,km\,s$^{-1}$}
\newcommand{\msol}{\,M$_{\odot}$}
\newcommand{\sfr}{\,M$_{\odot}$\,yr$^{-1}$}

\newcommand{\um}{\,$\upmu$m}

\title[Red quasars blow out molecular gas]{Red quasars blow out molecular gas from galaxies during the peak of cosmic star formation}

\author[H.\,R.~Stacey et al.]{H.\,R.~Stacey,$^{1}$\thanks{E-mail: stacey@mpa-garching.mpg.de}
T.~Costa,$^{1}$
J.\,P.~McKean,$^{2,3}$
C.\,E.~Sharon,$^{4}$
G.~Calistro~Rivera,$^{5}$
\newauthor E.\,Glikman $^{6}$
and P.\,P.~van~der~Werf $^{7}$
\smallskip \\ 
$^{1}$Max Planck Institute for Astrophysics, Karl-Schwarzschild Str. 1, D-85748 Garching bei M\"unchen, Germany \\
$^{2}$ASTRON, Netherlands Institute for Radio Astronomy, Oude Hoogeveensedijk 4, 7991 PD, Dwingeloo, The Netherlands \\
$^{3}$Kapteyn Astronomical Institute, University of Groningen, PO Box 800, 9700 AV Groningen, The Netherlands \\
$^{4}$Yale-NUS College, Singapore, 138527, Singapore \\
$^{5}$European Southern Observatory (ESO), Garching bei M\"unchen, Germany \\
$^{6}$Department of Physics, Middlebury College, Middlebury, VT 05753, USA \\
$^{7}$Leiden Observatory, Leiden University, PO Box 9513, NL-2300 RA Leiden, The Netherlands
}

\date{Accepted 2022 September 23. Received 2022 August 31; in original form 2022 July 18}

\pubyear{2022}

\begin{document}
\label{firstpage}
\pagerange{\pageref{firstpage}--\pageref{lastpage}}
\maketitle

\begin{abstract}
Recent studies have suggested that red quasars are a phase in quasar evolution when feedback from black hole accretion evacuates obscuring gas from the nucleus of the host galaxy. Here, we report a direct link between dust-reddening and molecular outflows in quasars at $z\sim2.5$. By examining the dynamics of warm molecular gas in the inner region of galaxies, we find evidence for outflows with velocities 500--1000\kms\ and timescales of $\approx0.1$~Myr that are due to ongoing quasar energy output. We infer outflows only in systems where quasar radiation pressure on dust in the vicinity of the black hole is sufficiently large to expel their obscuring gas column densities. This result is in agreement with theoretical models that predict radiative feedback regulates gas in the nuclear regions of galaxies and is a major driving mechanism of galactic-scale outflows of cold gas. Our findings suggest that radiative quasar feedback ejects star-forming gas from within nascent stellar bulges at velocities comparable to those seen on larger scales, and that molecules survive in outflows even from the most luminous quasars.
\end{abstract}




\section{Introduction}

State-of-the-art cosmological simulations and semi-analytic models of galaxy evolution invoke strong feedback from active galactic nuclei (AGN) in order to explain the observed properties of massive galaxies across cosmic time \citep{DiMatteo:2005,Somerville:2015,Schaye:2015, Weinberger:17, Dubois:21}. The injection of energy and momentum into the galactic interstellar medium (ISM) and circumgalactic medium by quasar winds and/or radio jets are predicted to regulate star formation in massive galaxies \citep{Fabian:2012}. This feedback is also thought to produce the observed scaling relations \citep{King:2005,Costa:2014} between galaxy properties (e.g. \citealt{Ferrarese:2000}) which already established by cosmic noon ($z\sim2\--3$; \citealt{ForsterSchreiber:2020}). However, the physical channels allowing energy and momentum released on sub-pc-scales to affect gas on galactic scales remain largely unconstrained from both theoretical and observational perspectives (\citealt{Harrison:2017,Harrison:2018}, for review). In particular, the question of whether AGN feedback operates predominantly via bulk ejection of the ISM or via prevention of cooling in the halo remains open. The answer to this question has profound implications on the nature of galaxy quenching and the ability of supermassive black holes to regulate their growth.

The most energetic class of AGN are quasars. Type~1 quasars are characterised by broad ionic or Balmer emission lines (FWHM $>1000$\kms). The majority of known quasars have unobscured optical continuum emission, although a subset (20--30~percent; \citealt{Glikman:2018b}) have been identified with reddened optical/infrared spectra. The nature of these reddened quasars has been a topic of debate for at least 20 years \citep{Webster:1995}. One might assume that they could fit into a scheme of AGN unification \citep{Antonucci:1993,Urry:1995} where the apparent quasar reddening is simply an effect of viewing angle, and broad line emission is observed but partially obscured by dust in the surrounding torus. However, many studies have found distinct properties of reddened quasars, indicating that they are intrinsically different from their bluer counterparts. Reddened quasars have significantly enhanced compact radio emission \citep{Klindt:2019,Rosario:2020,Fawcett:2020}, which may be linked to shocks from winds rather than the nucleus \citep{Hwang:2018,Rosario:2021}. Reddened quasars are also linked with higher-velocity outflows of ionised gas \citep{Urrutia:2009,Perrotta:2019,Temple:2019,Vayner:2021,Calistro:2021,Monadi:2022}, may have a different luminosity function \citep{Banerji:2015} and may have a higher incidence of mergers \citep{Glikman:2015,Zakamska:2019}. Furthermore, \cite{Calistro:2021} found no difference in torus column density in a large systematic study of the broad-band spectral energy distributions (SED) of red and blue quasars in SDSS. Altogether, this evidence strongly disfavours orientation as the primary explanation for reddening. The rarity of reddened quasars and the high incidence of outflows has led many studies to propose that they are caught in a short-lived `blow-out' phase in their evolution where rapid black hole accretion creates strong AGN feedback but has not yet cleared the nucleus of gas and dust (e.g. \citealt{Glikman:2012,Banerji:2015,Calistro:2021,Vayner:2021}).

An abundance of dust is conducive to strong quasar feedback. By enhancing the absorption cross section to UV and optical AGN radiation, the dusty ISM can, in principle, be driven out by radiation pressure, thereby reducing the supply of gas for star formation in the inner region of the galaxy \citep{Fabian:2008,Fabian:2009,Raimundo:2010,Ishibashi:2018} and resulting in correlations between galaxy bulge properties and black hole mass in agreement with observations \citep{Fabian:1999}. Radiation pressure on dust is predicted to operate and launch outflows at galactic scales on $100 \mathrm{pc} \-- 1 \mathrm{kpc}$ scales directly, as demonstrated by recent radiation-hydrodynamic simulations \citep{Bieri:17, Costa:2018a, Costa:2018b}. It may also launch fast winds from galactic torus scales ($\sim \rm pc$) \citep{Roth:12}, which shock-heat and generate hot, over-pressurised bubbles at larger scales \citep{Costa:2020}. Alternatively, ultra-fast winds with speeds $\sim 0.1 c$ can be driven out directly from accretion discs, though not via coupling to dust (which is expected to sublimate at those scales) but via radiation pressure on UV lines \citep[e.g.][]{Nomura:17} or magnetically \citep[e.g.][]{Fukumura:18}.

From an observational perspective, unobscured Type~1 quasars are ubiquitously associated with outflows of ionised gas (e.g. \citealt{Liu:2013,Rupke:2017}). However, the fuel for star formation is in the molecular phase, so observations of outflows of molecular gas can directly trace the impact of AGN feedback on the star formation. With the advent of the Atacama Large (sub-)Millimetre Array (ALMA), much attention has been payed to the search for molecular outflows from quasar host galaxies at high redshift. Most studies have focused on low-excitation CO emission or [CII], which probe the extent of the gas reservoir (e.g. \citealt{Sharon:2016,Neeleman:2021}), but evidence for widespread molecular outflows at high redshift ($z>1$) or higher AGN luminosities ($>10^{46}~{\rm ergs\,s^{-1}}$) is lacking. Some have found evidence for massive outflows of cold gas from quasar hosts on scales 10s--100s~kpc \citep{Bischetti:2019,Cicone:2021}, although this does not appear to be ubiquitous \citep{Novak:2020} and may be difficult to disentangle from bulk gas in the galaxy, mergers/companions and star-formation-driven outflows. It has been suggested that molecular gas is destroyed in the most energetic systems such that only ionised outflows are observed \citep{Fiore:2017}. 

To test this hypothesis, we have studied a sample of sixteen Type~1 quasar hosts at $z>1$ with carbon-monoxide (CO) molecular line measurements at rotational level transition $J_{\rm up}\geq7$. These lines probe warm, dense gas in the inner part of the galaxy ($\lesssim1$~kpc; \citealt{Stacey:2021}) that may be heated by the AGN radiation field \citep{Weiss:2007,vanderWerf:2010,Carniani:2019}. Here, we find distinct properties of high-J CO lines for red quasars that we attribute to outflow dynamics. In Section~\ref{section:data} we present the sample selection and introduce our new observations. In Section~\ref{section:analysis}, we explain the SED modelling and lens modelling of the lensed quasar systems. In Section~\ref{section:outflows} discuss the evidence for molecular outflows and the possible scenarios that might explain the differences in CO properties. In Section~\ref{section:mechanisms}, we discuss the quasar feedback mechanisms that could power the molecular outflows. In Section~\ref{section:conclusion}, we consider the implications for our results for galaxy and quasar evolution, and discuss avenues for future work. Throughout, we assume the \cite{Planck:2016} flat $\Lambda$CDM cosmology with $H_{0}= 67.8~$km\,s$^{-1}$ Mpc$^{-1}$, $\Omega_{\rm M}=0.31$ and $\Omega_{\Lambda}=0.69$.

\section{Observations}
\label{section:data}

For our analysis, we searched the literature and the ALMA archive for observations of Type~1 quasars in $J\geq7$ CO lines. All these observations have sub-arcsecond angular resolution to account for any companions that may contaminate our measurements. Most of these quasars are strongly-lensed and were discovered in optical/infrared surveys.

\begin{table*}
    \caption{New ALMA and NOEMA observations presented in this work. We give the project code, the synthesised beam FWHM ($\Theta_{\rm beam}$) used to extract the line profile and continuum flux density, the continuum frequency ($\nu_{\rm cont}$), the continuum flux density ($S_\nu$) and the integrated line intensity ($I_{\rm line}$) based on a Gaussian fit to the line profile(s). ${\dagger}$ denotes cases where we have applied a $uv$ taper to the data weights to improve surface brightness sensitivity. $^*$For these cases, the line intensity is approximated from the Gaussian fit to the line profile (see Section~\ref{section:data}).}
    \centering
    \setlength{\tabcolsep}{5pt}
    \begin{tabular}{l l c c c c c c c} \hline 
        Project & Name & Line(s) & $\Theta_{\rm beam}$ & $\nu_{\rm cont}$ & $S_{\nu}$ & $I_{\rm line}$ \\ 
        & & & (arcsec) & (GHz) & (mJy) & (Jy\kms) & \\ \hline 
        S17BV & MG\,J0414+0534 & CO\,(3--2) & $7.8\times2 .7$ & 96 & $34.8\pm0.5$ & $3.1\pm0.4^{*}$ \smallskip\\ 
        2018.1.01008.S & MG\,J0414+0534 & CO\,(7--6); [CI]\,(2--1) & $0.3\times0.3^{\dagger}$ & 236 & $10.1\pm0.4$ & 7.8$^{*}$; 3.7$^{*}$ \smallskip\\ 
        2019.1.00948.S & SDSS\,J1330$+$1810 & CO\,(7--6); [CI]\,(2--1) & $0.5\times0.4$ & 342 & $6.3\pm0.2$ & $3.5\pm0.2$; 3.9$^{*}$ \smallskip\\
        2019.1.00964.S & DES\,J0408$-$5354 & CO\,(7--6); [CI]\,(2--1)  & $1.2\times1.1^{\dagger}$ & 247 & $3.0\pm0.4$ & $4.9\pm0.7$; $3.3\pm0.5$ \\
        & J1042$+$1641 & CO\,(10--9) & $0.9\times0.7$ & 256 & $2.0\pm0.3$ & $4.0\pm0.7$ \\ \hline 
        \end{tabular} 
    \label{table:alma_obs}
\end{table*}

\begin{figure*}
    \centering
    \includegraphics[height=0.22\textheight]{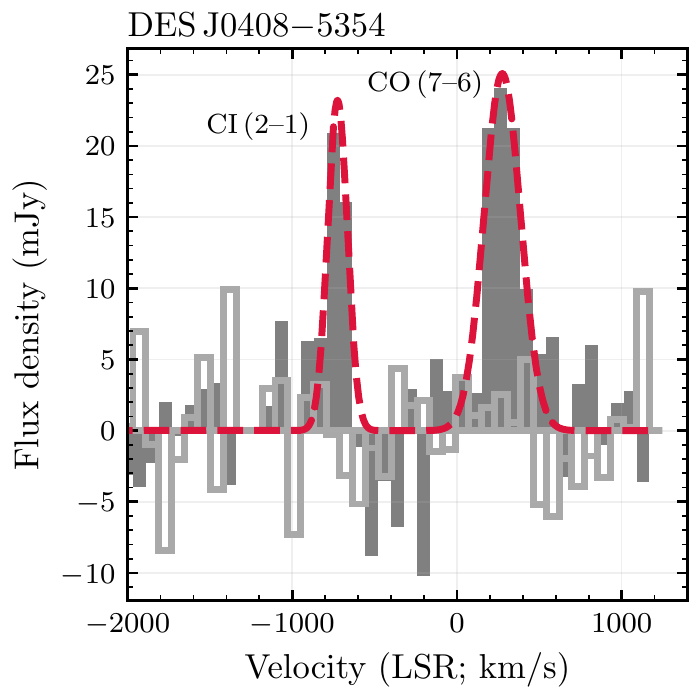}
    \includegraphics[height=0.22\textheight]{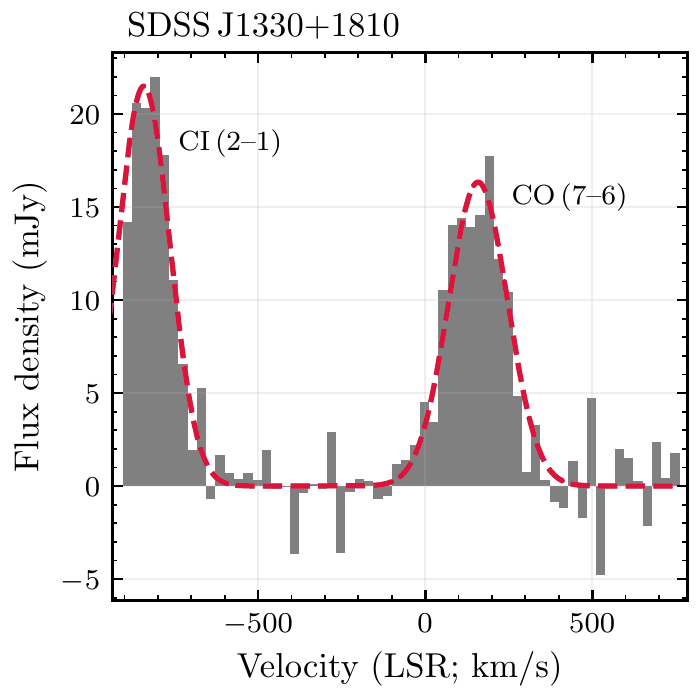}
    \includegraphics[height=0.22\textheight]{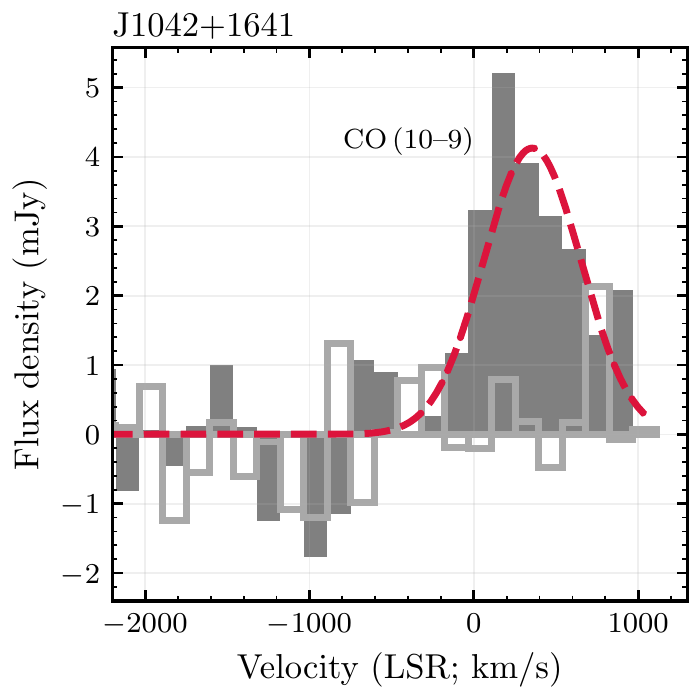}
    \includegraphics[height=0.22\textheight]{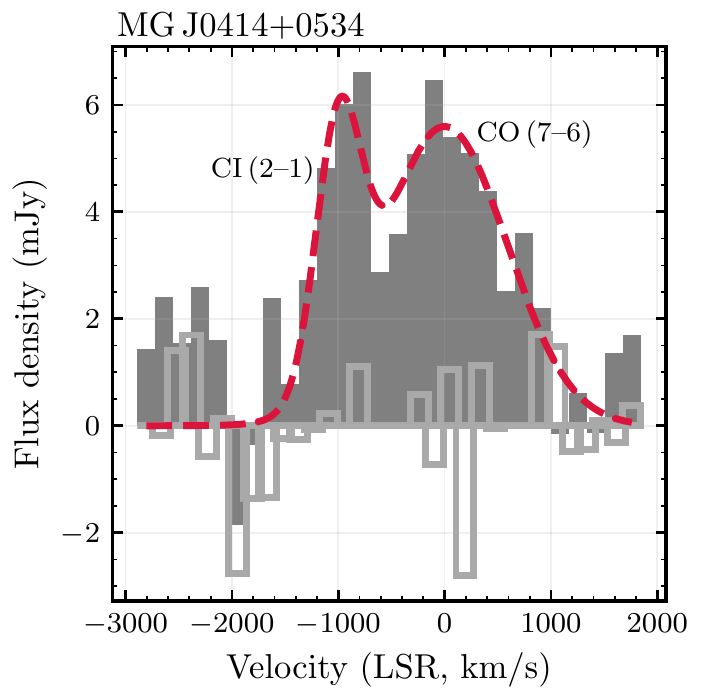}
    \includegraphics[height=0.22\textheight]{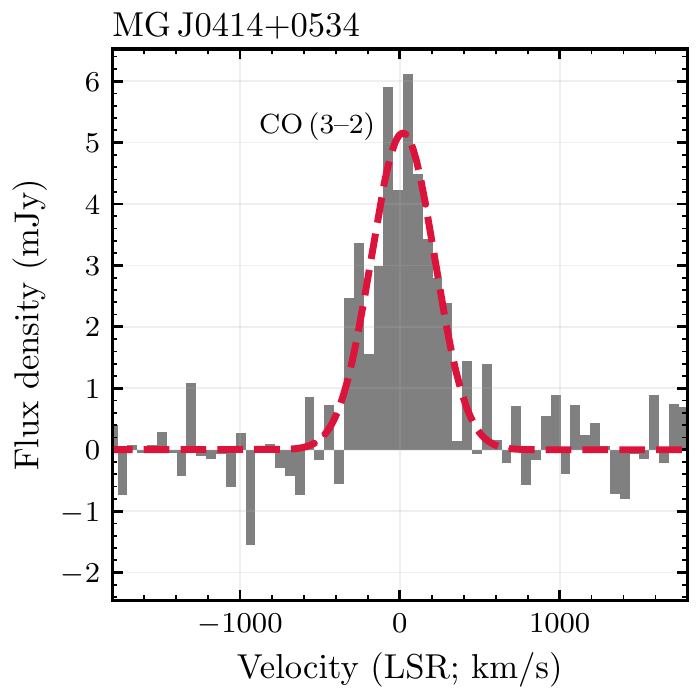}
    \caption{Line profiles for the new ALMA and NOEMA data presented in this work. The spatially-integrated line profiles (solid grey) and Gaussian fits to the CO and [CI]\,(2--1) line emission (red). Noise in the adjacent spectral window is shown in light grey unfilled bars for the noisier spectra. For all except MG~J0414+0534 (bottom row), the systemic redshift is based on optical spectroscopy so a velocity offset in the CO is most likely because the ionic/Balmer lines trace out-flowing gas.}
    \label{fig:line_profiles}
\end{figure*}

\begin{figure}
    \includegraphics[width=0.46\textwidth]{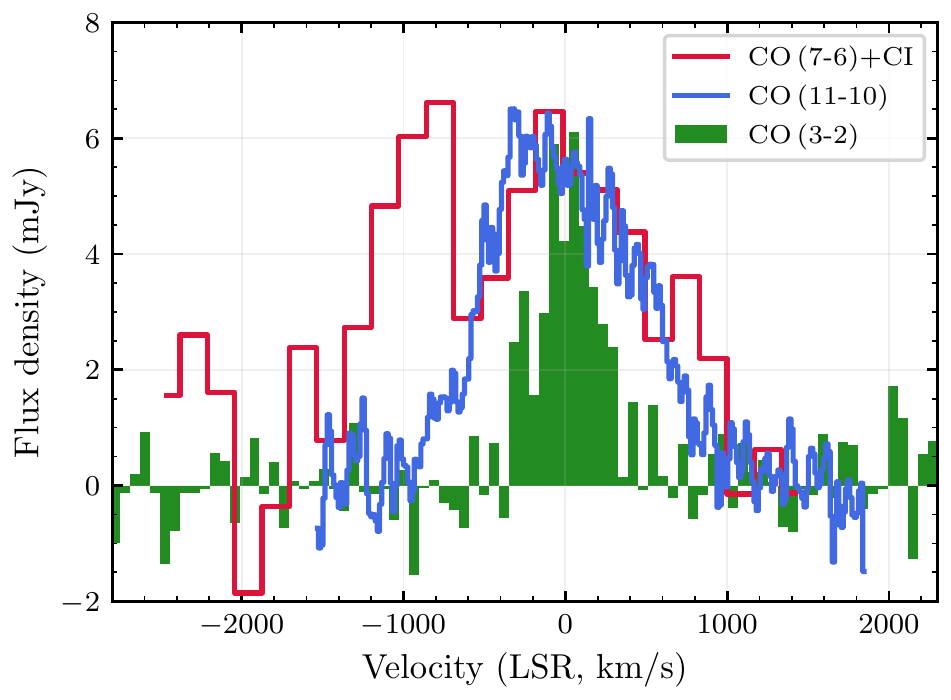}
    \vspace*{-0.2cm}
    \caption{Comparison of CO line profiles for MG\,J0414+0534. While the CO\,(7--6) and [CI]\,(2--1) lines are blended, the CO\,(7--6) is similar to the CO\,(11--10) while both are clearly broader than the CO\,(3--2). This suggests that the CO\,(7--6) and CO\,(11--10) both probe different dynamics than the CO\,(3--2). The systemic redshifts of these lines are all consistent.}
    \label{fig:0414_CO_lines}
\end{figure}

\subsection{ALMA observations}

In addition to CO, [CI] and [CII] measurements obtained from the literature, we present new ALMA data for several objects. These observations are summarised in Table~\ref{table:alma_obs}. Where we observe the CO~(7--6) lines we also observe the [CI]\,(2--1) line simultaneously. Each data set was calibrated and reduced using the ALMA pipeline in CASA \citep{McMullin:2007}. The calibrated data were inspected to confirm the quality of the pipeline calibration. We self-calibrated the data for SDSS\,J1330$+$1810 using the continuum-only spectral windows with a single solution interval for each antenna in both amplitude and phase. For MG~J0414$+$0534, we performed phase calibration with a solution interval of 240~seconds.

We created a clean image cube of the spectral window containing the line emission with natural weighting of the visibilities (images are shown in Fig.~\ref{fig:images} of the Appendix) and extracted a spectrum in an aperture around the lens. The line profiles (presented here and in the literature) do not show asymmetries that would indicate that their shapes are strongly affected by differential magnification (Fig.~\ref{fig:line_profiles}). J1042+1641 is a possible exception, although this may be explained by the signal-to-noise ratio of the data.

To measure the line width, we fit the line profiles with a single Gaussian, taking into account the noise in each channel. For MG~J0414$+$0534, the CO~(7--6) line is so broad that it is blended with the [CI]~(2--1) line. We fit Gaussians to both lines simultaneously with the systemic redshift fixed to the value found for the molecular gas \citep{Barvainis:1998,Stacey:2020}. For SDSS\,J1330$+$1810, part of the [CI] line falls outside the spectral window. For these two cases, the integrated line luminosity is estimated from the fitted Gaussian profile rather than directly from the image. The line profiles and Gaussian fits for these ALMA observations are shown in Fig.~\ref{fig:line_profiles}.

\subsection{NOEMA observations}

We present new data from the WideX correlator on the Northern Extended Millimetre Array (NOEMA) of CO\,(3--2) line emission from MG\,J0414+0534. Data calibration, flagging for data quality, and imaging were carried out using the IRAM GILDAS package \citep{Guilloteau:2000}. The standard calibration pipeline was lightly modified to compensate for poor weather, low elevation observations, and source brightness. The modifications include relaxing data flagging conditions (with some manual flagging as necessary), applying a separate phase correction based on water vapour radiometer observations over a longer scan period, and self-calibration using the line-free channels of MG\,J0414+0534. Due to the high signal-to-noise ratio of the observations, self-calibration required two model regions: an elliptical Gaussian and a point source, corresponding to the South-East image pair and Northern image, respectively, as seen in prior higher angular resolution images of this gravitationally lensed system.

To create a continuum-free image cube, we performed continuum subtraction in Fourier space using standard GILDAS routines. Due to substantial side lobes in the synthesised (dirty) beam caused by the low-elevation/short-duration observations, we used a custom beam fitting script to determine the size of the central peak for the clean beam. The spatially integrated line profile was extracted in the same method as for the ALMA observations and the line width and integrated line flux was measured by fitting a single Gaussian.

\section{Model fitting}
\label{section:analysis}

\subsection{Lens modelling}

Reconstruction of lensed emission may be performed assuming parametric sources \citep{Spilker:2016} or pixellated sources \citep{Stacey:2021}, depending on the science goal and data quality. We choose a parametric source as we are interested in extracting a measure of the intrinsic source size and the data quality is generally not sufficient to allow for more freedom in the source surface brightness distribution.

We perform lens modelling for ALMA data using the software {\sc visilens} \citep{Spilker:2016} that fits the visibility data directly to avoid errors introduced by deconvolution and sparse coverage of the Fourier plane. The model parameters are inferred using a Markov chain Monte Carlo method. We adopt fiducial lens models from the literature consisting of a singular isothermal ellipsoid and external shear, and in some cases an additional singular isothermal sphere for a companion galaxy. We model the source as a Gaussian or several Gaussians. The details of the fits to the continuum and integrated line emission are given in Table~\ref{table:lensmodels} of the Appendix. Images of the maximum a-posteriori lens models are shown in Figs.~\ref{fig:0408_cont}--\ref{fig:1330_CI} of the Appendix. 

For MG\,J0414+0534, approximately half of the continuum emission at 350~GHz is due to synchrotron emission from the AGN \citep{Stacey:2018b}. This complicates the lens modelling as the compact emission is thought to be affected by dark matter structure \citep{Stacey:2018b}. We attempt to overcome this by making a deconvolved image with superuniform weighting (0.1~arcsec resolution) where the compact, high-brightness-temperature emission dominates the signal. We subtracted these CLEAN components from the visibility data such that the remaining emission is assumed to be only the dust continuum. This is not a perfect decomposition of the source emission mechanisms, but sufficient to estimate the magnification of the dust. 

\subsection{AGN luminosity, extinction and black hole masses}

We use {\sc AGNFitter} \citep{Calistro:2016} to fit the broad-band SED of each quasar and host galaxy. This software includes accretion disc, torus, host galaxy and starburst templates \citep{Calistro:2016,Calistro:2021}. The accretion disc component is reddened by an SMC extinction dust law \citep{Prevot:1984} parameterised by rest-frame $E_{B-V}$. We compile photometry from all-sky surveys, where available, or from the literature.

We assume a minimum uncertainty of 10~percent on photometric measurements to allow for source variability, microlensing-induced variability and systematic calibration offsets. Although the photometry will be blended with light from the lensing galaxy, the quasar point-source emission strongly dominates the emission. This is evident from the available ultra-violet--infrared imaging with the Hubble Space Telescope (e.g. \citealt{Kochanek:1999}). The host galaxy emission can be fit in only a few cases, for most only accretion disc and torus components are constrained. We find some minor underfitting which may be due to intrinsic reddening of the empirical templates (e.g. RX\,J0911+0551) or microlensing of the lensed quasar emission.

Following \cite{Calistro:2021}, we calculate the bolometric AGN luminosity ($L_{\rm AGN}$) by integrating the quasar accretion disc model (`big blue bump') in the range 0.05--1\um\,with an additive correction of 0.3\,dex to account for X-ray emission not included in the model fit, i.e.
\begin{equation}
    \log L_{\rm AGN} = \log L_{\rm 0.05\--1\upmu m} + 0.3 - \log\mu_{\rm qso} ,
\end{equation} where $\mu_{\rm qso}$ is the quasar magnification. We assume black hole mass ($M_{\rm BH}$) estimates from the literature that were derived using the virial method. For consistency, we adopt the same quasar lensing magnification used for the black hole mass to infer the intrinsic bolometric AGN luminosity. For SDSS\,J1330+1810, the ionic or Balmer line widths have not been reported, so we infer a lower limit the black hole mass from the AGN bolometric luminosity assuming Eddington-limited accretion.

Plots of the IR--UV SED fitting is shown in Fig.~\ref{fig:SEDs} of the Appendix. 

\subsection{Star formation rates}

Where possible, we obtain far-infrared luminosities from the literature. While {\sc AGNFitter} includes templates to fit the host galaxy dust emission, these models are empirically derived from low-redshift starbursts and may not be appropriate for high-redshift quasar hosts. Indeed, we found that the far-infrared--mm spectra are often poorly fit by these templates. Instead, we adopt far-infrared luminosities taken from the literature, most of which were calculated with modified black-body models, where the effective dust temperature and dust emissivity may be free parameters. Where there is insufficient photometry to fit a dust model, we assume an optically-thin modified black-body with the median effective dust temperature and emissivity found for a large sample of quasar hosts \citep{Stacey:2018a}. While the effective dust temperature may be quite different depending on the choice of model, the integrated luminosity will be similar as it is constrained by the photometry.

We convert far-infrared luminosity (40--120\um) to total infrared luminosity (8--1000\um) using the colour-correction factor \citep{Dale:2001} of 1.91 and to a dust-obscured SFR using a standard calibration \citep{Kennicutt:1998} assuming a Salpeter initial mass function. While it has been proposed that the SFRs of AGN-starbursts may be overestimated due to an unconstrained contribution from AGN-heated dust, radiative transfer modelling of far-infrared lines from the most star-forming objects in the sample still support the existence of high star formation rate densities in combination with AGN heating \citep{vanderWerf:2011,Li:2020,Wang:2019,Uzgil:2016}.

The intrinsic SFR for some of the lensed sources have previously been inferred from lens modelling of ALMA observations of sub-mm dust emission \citep{Stacey:2021}. For APM\,08279$+$5255, we adopt the CO magnification factor from lens modelling in the literature to infer a SFR from the uncorrected measurement \citep{Riechers:2009}. All values and references are given in Table~\ref{table:sources} of the Appendix.

Note that the sub-mm magnification is not the same as the quasar magnification, as these will be lensed differently depending on the size and location of the emitting region. We did not check for consistency between the lens models used to obtain black hole masses and those used to obtain the SFR, but these two properties are not directly compared here.


\begin{figure*}
    \centering
    \includegraphics[width=0.72\textwidth]{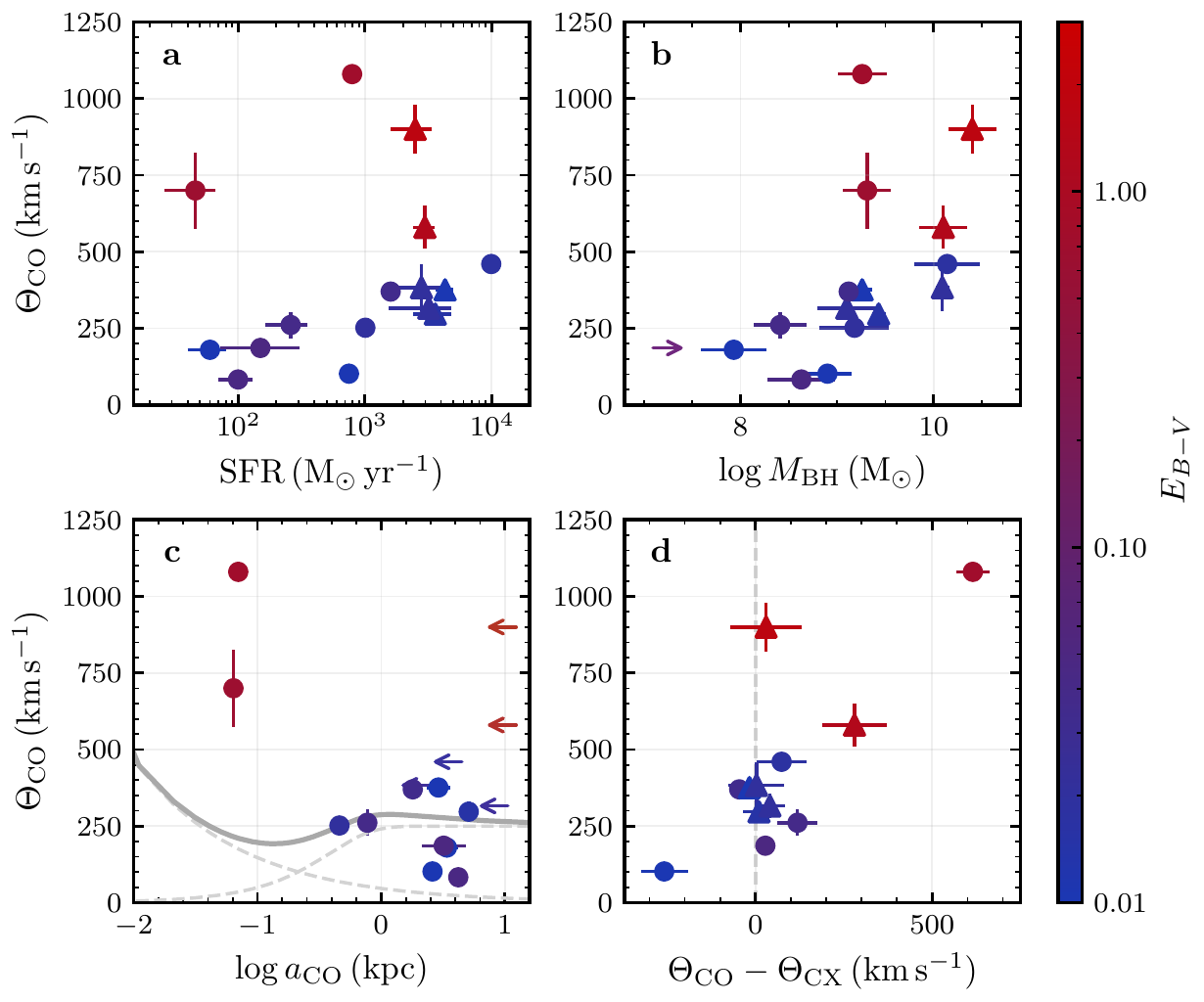}
    \vspace{-0.2cm}
    \caption{Comparison of CO line widths of the sample with reddening and various galaxy properties. \textbf{a}: $J_{\rm up}>7$ CO line width against SFR (lensing-corrected). \textbf{b}: CO line width against $M_{\rm BH}$ (lensing-corrected). \textbf{c}: FWHM of the physical size\protect\footnotemark of the CO emission ($a_{\rm CO}$, i.e. twice the effective radius) against CO velocity line width. Twice the effective radius is often assumed to calculate dynamical mass of a disc \citep{ForsterSchreiber:2018,Neeleman:2021}. The grey curves show a Keplerian rotation curve for a $10^{9.3}$\msol\ black hole (the black hole masses of the two red quasars shown in this figure) and a rotation curve with peak velocity 500\kms\ and transition radius of 500~pc adopted from  fast-rotating massive galaxies at $z\simeq4.5$ \citep{Fraternali:2021}, both inclined at 45~deg. \textbf{d}: The difference between the line width of the high-J CO line and the line width of a bulk gas tracer ([CII], CO\,(3--2) or [CI]). The circles identify the sources that are lensed. The colour scale indicates quasar extinction ($E_{\rm B-V}$). The red quasars inhabit a different parameter space than the blue quasars across all indicators, while the blue quasars follow expected trends with dynamical mass and/or stellar feedback. }
    \label{fig:FWHM_SFR}
\end{figure*}\footnotetext{If a system is modelling by multiple Gaussian sources, the size is a flux-weighted average}

\section{Evidence for molecular outflows}
\label{section:outflows}

Details of the line properties, SFR, $M_{\rm BH}$, $E_{\rm B-V}$, $L_{\rm 0.05\--1\upmu m}$, adopted quasar magnifications and literature references for the sample of quasars is presented in Table~\ref{table:sources} of the Appendix. The sample probes 2, 3 and 4 orders of magnitude in SFR, black hole mass and AGN luminosity, respectively.  

The sample consists of a range of CO rotational transitions, $J_{\rm up}=7$--11. Previous studies of CO lines from quasar hosts have found generally consistent line widths for these high-excitation CO transitions relative to mid-J and low-J CO lines \citep{Weiss:2007,Li:2020}, although their kinematics have not yet been systematically studied at high redshift so little is previously known about their typical dynamics. We compare the CO\,(7--6) and CO\,(11--10)\footnote[2]{The CO\,(11--10) line is used for MG\,J0414+0534 in all plots} line widths for MG~J0414$+$0534 and find comparably large values of $\approx1000$\kms. As shown in Fig.~\ref{fig:0414_CO_lines}, these lines are significantly broader than the CO\,(3--2) line from the same system, suggesting that the CO\,(7--6) and CO\,(11--10) both probe dynamics that are different from the CO\,(3--2).

For the blue quasars, we find positive correlations between CO line width and SFR ($p=0.002$) and between CO line width and black hole mass ($p=0.03$) with the Pearson correlation test. Such correlations can be explained by well-understood physical phenomena. The correlation with SFR (Fig.~\ref{fig:FWHM_SFR}a) could reflect increased turbulence due to radiative stellar winds or supernovae: gas dispersion of up to 100\kms\, can be induced by stellar feedback in regions of Eddington-limited star formation \citep{Narayanan:2014,Hung:2019} (note that if the SFRs are overestimated due to a contribution from AGN-heating \citep{Kirkpatrick:2015}, the induced turbulence may be lower). Alternatively, it may reflect the relationship between SFR and dynamical mass. The correlation between line width and black hole mass (Fig.~\ref{fig:FWHM_SFR}b) may reflect a larger dynamical mass as expected from canonical scaling relations between black hole mass and stellar mass \citep{Ferrarese:2000}. These are expected correlations found for [CII] in quasar host galaxies \citep{Neeleman:2021}, although these correlations have not been previously investigated for high-excitation CO lines. The red quasars do not follow the correlations observed by the blue quasars and there appears to be no significant correlation between SFR or black hole mass and reddening ($p=0.45$ and $p=0.24$, respectively, for the whole sample) in agreement with previous work \citep{Calistro:2021}, although the red quasars in the sample have larger black holes on average. 

In Fig.~\ref{fig:FWHM_SFR}c, we compare the resolved size of the high-J CO line emission to the line width to determine whether these velocities could be produced by the dynamical mass of the galaxy. We find that the size of the emission from the blue quasars is $\sim1$~kpc, consistent with gas that could be from within a massive disc. For the two red quasars with resolved high-J CO line emission, the sizes of $\approx70$~pc would suggest an unfeasibly large enclosed mass of $\sim10^{10}$\msol. This could be reconciled for J1042+1641 if the CO emission was from a Keplerian disc seen edge-on, although this is not consistent with the Gaussian shape of the line profile.

We also compare the high-excitation CO with [CII], CO\,(3--2) or [CI] line profiles to further test whether the line is associated with the bulk gas dynamics. [CII], CO\,(3--2), [CI]\,(1--0) and [CI]\,(2--1) trace the bulk of gas at lower temperatures and densities, so they are expected to probe a galaxy at larger disc radii (i.e. maximum circular velocity; e.g. \citealt{Banerji:2021,Fraternali:2021}). Fig.~\ref{fig:FWHM_SFR}d shows the CO line width compared to the difference between the CO line width and the line width of a bulk gas tracer\footnote{ Our order of preference is [CII], CO\,(3--2) and [CI]~(2--1), based on the relative prevalence of previous studies of the dynamics of the lines. }, for the sources with such measurements. The difference is around zero or less for the blue quasars, so these lines could both relate to gas in the host galaxy. However, two out of three red quasars have CO line widths larger than their bulk gas line widths. This is not a clean test of the high-J CO dynamics as the kinematics and physical conditions of [CII], CO\,(3--2) and [CI]~(2--1) lines are not systematically studied for quasar hosts (in particular, [CI]~(2--1) may also be produced by cosmic ray excitation of molecular gas by AGN radiation), but larger velocities of high-J CO lines is difficult to explain from a radiative transfer perspective if they are co-spatial.

Both the lack of relationship with SFR and the compact size of the CO emission that is coincident with the quasar disfavour these being star-formation-driven outflows, and are contrary to findings of outflows from high-redshift star-forming galaxies \citep{Ginolfi:2020}. Furthermore, if the broad high-J CO lines are driven by intense star formation, we should also expect to see outflows from blue quasars which are hosted in galaxies with Eddington-limited star formation rate densities, which we do not.

In summary, we conclude that quasar-driven outflows are the most likely explanation for the high-J CO linewidths of red quasars based on the following of evidence:
\begin{enumerate}\vspace{-2pt}
  \setlength\itemsep{2pt}
    \item the lack of relationship with star formation rate;
    \item the lack of relationship with black hole mass (stellar mass);
    \item the larger velocities than lines that commonly probe the bulk of the gas in the galaxy for 2/3 red quasars; and
    \item the sizes of the line emission for the two red quasars where the high-J CO can be resolved that imply an unfeasibly large dynamical mass.
\end{enumerate}\vspace{-2pt}

\begin{figure}
    \includegraphics[width=0.45\textwidth]{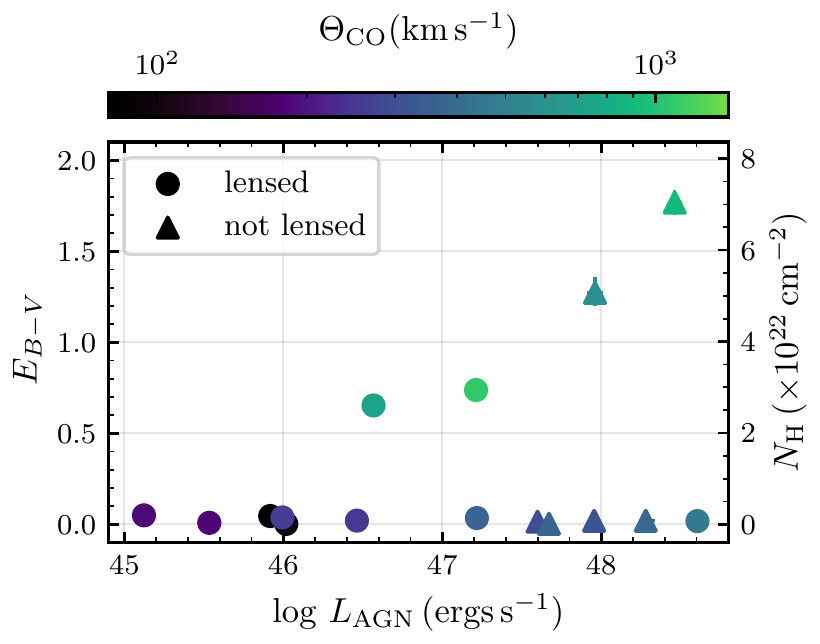}
    \vspace{-0.1cm}
    \caption{Reddening against bolometric AGN luminosity (lensing-corrected) coloured by high-J CO line width. The opposing y-axis shows the corresponding column density \protect\citep{Genzel:2013}. Circles identify sources that are lensed.}
    \label{fig:FWHM_LAGN}
\end{figure}

Using an orthogonal distance regression, we fit a linear relationship between CO line width and SFR for the blue quasars, as this shows a statistically significant correlation and the contribution of outflows to the lines tracing the bulk gas for individual red quasars is unknown. Assuming the line profiles have contributions from both outflows and host galaxy that can be described by Gaussians of equal height, we estimate line-of-sight outflow velocities of 500--1000\kms. These velocities may be intrinsically larger depending on the geometry of the outflows. More accurate estimates of the outflow velocities will require detailed follow-up observations.

As we show in Figs.~\ref{fig:FWHM_LAGN} and \ref{fig:FWHM_LEdd}, there is no indication that the red quasars in our sample have larger AGN luminosities or larger Eddington ratios than the blue quasars, consistent with previous work by \cite{Calistro:2021} but in contrast to \cite{Kim:2015}.

\section{Outflow driving mechanism}
\label{section:mechanisms}

\subsection{Radiation pressure}

Recent radiation-hydrodynamic simulations \citep{Bieri:17,Costa:2018a,Costa:2018b} predict that radiation pressure operates and launches outflows from $<1$~kpc scales,  thereby reducing the supply of gas for star formation in the inner region of the galaxy. This blow-out phase is likely short-lived because outflows propagate on a characteristic timescale of 
\begin{equation}
    t_{\rm out} \sim  1 \left( R_{\rm out} / \rm{kpc} \right) \left( v_{\rm out} / 1000 \, \rm{km \, s^{-1}}\right)^{-1} \, {\rm Myr} .
\label{eq:timescale}
\end{equation} 
For this reason, we expect quasars to be below the effective Eddington limit unless their host galaxies are experiencing a powerful outflow. This limit creates a `forbidden zone' on the column density ($N_{\rm H}$) -- Eddington ratio ($\lambda_{\rm Edd}$) plane in the regime where the dust is optically thick to ultra-violet radiation which can approximated by
\begin{equation}
    N_{\rm H} \sim \frac{\lambda_{\rm Edd}/\sigma_{\rm T}}{1 - \frac{\kappa_{\rm IR}}{\kappa_{\rm T}}\lambda_{\rm Edd}},
    \label{eq:effective_eddington}
\end{equation} where $\sigma_{\rm T}$ is the Thompson cross-section, $\kappa_{\rm T}$ is the electron scattering opacity and $\kappa_{\rm IR}$ is the dust opacity \citep{Ishibashi:2018}. Previous studies \citep{Fabian:2009,Ishibashi:2018} have defined lower limits for the forbidden zone of $10^{21.5}$--$10^{22}\,{\rm cm^{-2}}$ where column densities could be associated with diffuse cold gas in the galaxy: we have adopted the more conservative assumption of $10^{22}\,{\rm cm^{-2}}$, although the choice does not affect our interpretation. 

\begin{figure}
    \centering
    \includegraphics[width=0.48\textwidth]{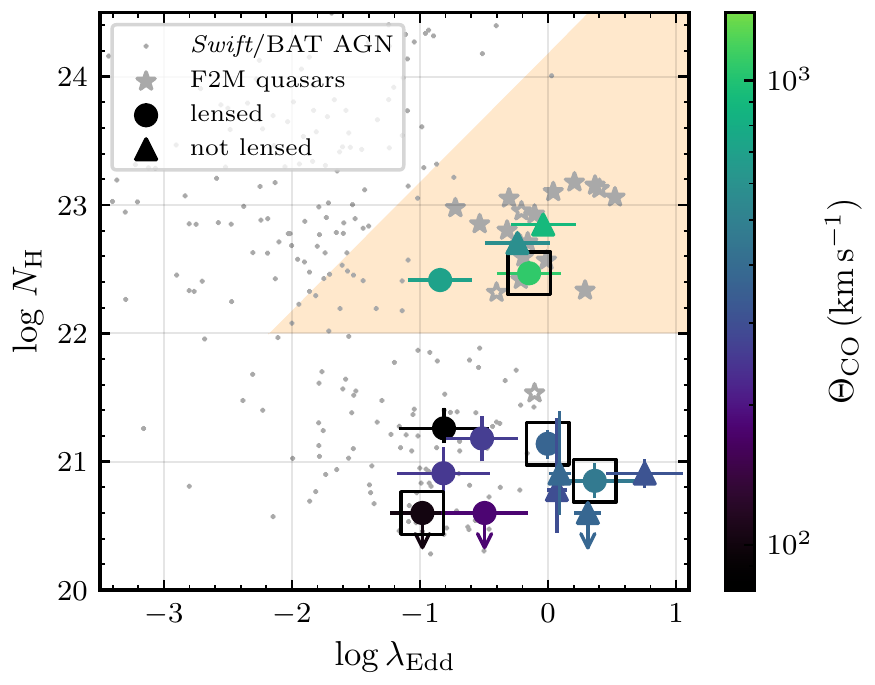}
    \caption{Quasar sample relative to the forbidden zone where radiation pressure on dust is sufficient to expel the obscuring column density. The shaded region identifies the `forbidden zone' where quasars are expected to be in a short-lived blow-out phase \protect\citep{Ishibashi:2018}. The red quasars are in the forbidden zone where radiation pressure on dust can drive out the obscuring gas, indicating that radiation pressure is responsible for the outflows. Coloured symbols indicate the objects in this work, coloured according to their $J\geq7$ CO line width. Boxes identify the objects in our sample with UV/X-ray ultra-fast outflows or broad absorption lines \protect\citep{Chartas:2000}. The stars show quasars from F2M surveys ($0.1<E_{\rm B-V}<2.5$; \protect\citealt{Glikman:2012}) where $N_{\rm H}$ is inferred from $E_{\rm B-V}$ (filled stars) or X-ray spectra (open stars) \protect\citep{Glikman:2017}. The dots are X-ray detected AGN in the {\it Swift}/BAT sample \protect\citep{Ricci:2017}, where $N_{\rm H}$ is inferred from X-ray spectra.}
    \label{fig:FWHM_LEdd}
\end{figure}

\begin{figure}
    \includegraphics[width=0.47\textwidth]{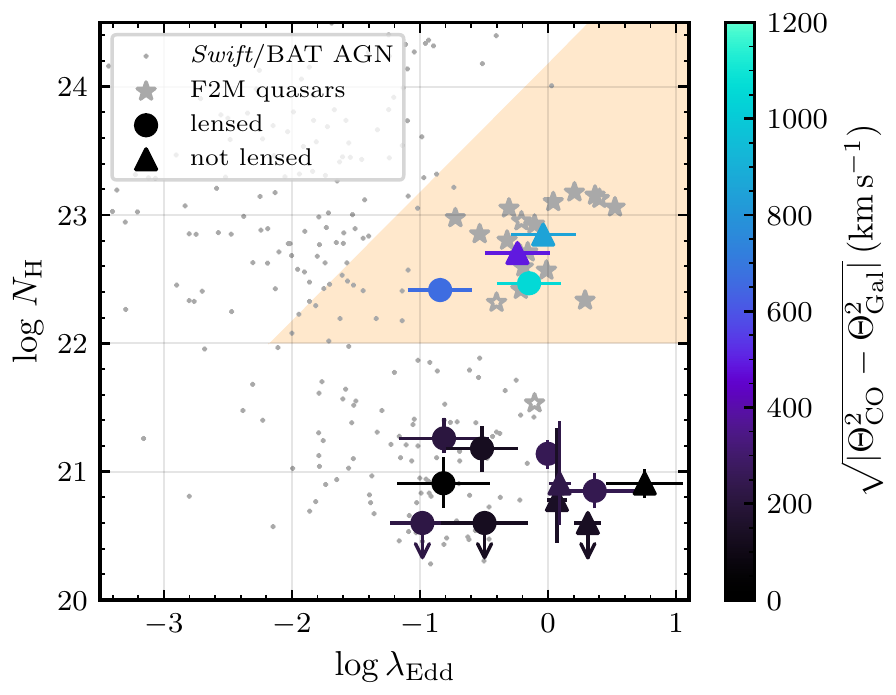}
    \caption{Quasar sample relative to the forbidden zone where radiation pressure on dust is sufficient to expel the obscuring column density. Same as Fig.~\ref{fig:FWHM_LEdd} but with the colours indicating the estimated outflow contribution (see Section~\ref{section:outflows}).}
    \label{fig:FWHM_LEdd_outflow}
\end{figure}

We convert the $E_{\rm B-V}$ from our SED fitting to a gas column density using the relationship found for a $z=1.5$ star-forming galaxy by \cite{Genzel:2013}, assuming that the molecular gas column density is a good approximation for the total gas column density. Our SED fitting uncertainties suggest we cannot constrain $E_{\rm B-V}$ at values below 0.01, so adopt this as an upper limit for the equivalent column density. As shown in Figs.~\ref{fig:FWHM_LEdd} and \ref{fig:FWHM_LEdd_outflow}, all four red quasars lie in the forbidden zone, suggesting that radiation pressure is responsible for these outflows. In contrast, the blue quasars populate a region of the parameter space where the nucleus is unobscured suggesting that they have already driven out their surrounding gas, consistent with the lack of a high velocity molecular gas component.

Two red quasars in our sample have measured sizes of their CO emission from lens modelling in this work (Figs.~\ref{fig:0414_cont} and \ref{fig:1042_line}). From Eq.~\ref{eq:timescale} and the estimated outflow contribution, we estimate an outflow timescale of $\approx0.1$~Myr for MG\,J0414+0534 and J1042+1641. This is within the expected timescale of a radiatively-driven blow-out \citep{Ishibashi:2017,Ishibashi:2018}.

\subsection{Testing an energy-driven scenario}

Fig.~\ref{fig:FWHM_LEdd} supports the scenario that radiation pressure drives the observed outflows. However, outflows may also be `energy-driven' by high-velocity winds produced near the accretion disc \citep{King:2010,Faucher-Giguere:2012,Costa:2014,Costa:2020}. Ultra-fast outflows have been detected in X-ray spectra of high-redshift quasars \citep{Chartas:2021} including the red quasar MG\,J0414+0534, which contains an ultra-fast outflow with velocity 0.3$c$ \citep{Dadina:2018} that could potentially power the observed molecular outflow. The kinematics of the wind is a key factor to discriminate between the two scenarios: the momentum flux for an energy-driven outflow is expected to be
\begin{equation}
    \dot{p_{\rm out}} \sim \frac{v_{\rm UFO}}{v_{\rm out}} \frac{L_{\rm AGN}}{c},
\end{equation} where $v_{\rm out}$ and $v_{\rm UFO}$ are the velocities of the molecular outflow and ultra-fast outflow, respectively \citep{Costa:2018b}. For the inferred $L_{\rm AGN}=10^{47.2}~{\rm ergs\,s^{-1}}$, we predict an energy-driven mass outflow rate ($\dot{M} \, = \, \dot{p} / v_{\rm out}$) of $\approx10^{6}$\sfr\,for the gas probed by the CO\,(11--10). Adopting a simple spherical model for the molecular outflow with radius of 70~pc and velocity 1000\kms, the implied outflow timescale is $\approx0.07$~Myr and thus the total outflowing mass is $\approx10^{10}$\msol. The CO\,(1--0) line is commonly used to probe the bulk molecular gas mass in galaxies in the high-redshift Universe, assuming a CO-to-H$_{2}$ conversion factor: adopting the 3$\sigma$ upper limit for the non-detection of CO\,(1--0) \citep{Sharon:2016}, a typical \citep{Greve:2014} conversion factor of $0.8\,{\rm K\,km\,s^{-1}\,pc^{2}}$ and magnification factor of 10 \citep{Stacey:2018a}, we estimate the molecular gas mass in the host galaxy is $<10^{9.5}$\msol\footnote{Note that this would be a factor of 2 lower if we assume the dust magnification in Table~\ref{table:lensmodels}}. Therefore, for an energy-driven wind model, the gas in the outflow from MG\,J0414+0534 would be more than twice as massive as the molecular gas in the disc: an unlikely scenario given the high SFR (Table~\ref{table:sources}) and the massive molecular gas reservoirs of quasar hosts that are similar to normal starbursts \citep{Riechers:2011a,Riechers:2011b,Sharon:2016}. This disfavours an energy-driven scenario for this object, requiring that an accretion disc wind must couple very inefficiently with the ISM in order to produce the outflow we observe. 

In Fig.~\ref{fig:FWHM_LEdd}, we identify the quasars in our sample where ultra-fast outflows or broad absorption lines have been detected in UV or X-ray spectra. Three out of four quasars that have such features do not show evidence for high-velocity molecular gas, suggesting that radiation pressure on dust is a better predictor of the presence of molecular outflows. 

\section{Conclusions}
\label{section:conclusion}

It has frequently been suggested that strong quasar reddening is related to a transitional `blow-out' phase in which quasars evacuate the obscuring gas in the nuclear region (e.g. \citealt{Glikman:2012,Glikman:2017,Banerji:2012,Calistro:2021}). Here, we find evidence for a direct connection between quasar dust-obscuration and molecular outflows. This phenomenon is predicted by theoretical models of radiation pressure on dust and is in agreement with observational evidence for fundamentally different properties of red quasars. Our results indicate that a radiatively-driven blow-out could be a viable mechanism to produce the rapid quenching of star formation \citep{Belli:2021,Williams:2021} and depleted molecular gas reservoirs \citep{Whitaker:2021} recently observed in quiescent galaxies at cosmic noon.

In addition, we introduce a new approach to identifying molecular outflows through high-excitation CO lines. Previously, much sought-after cold outflows have typically been identified from offsets from the systemic velocity (which is often ill-defined for quasars) and broad wings of [CII] lines (which may be atomic or ionised gas), while we find neither here. The dearth of detection of cold outflows has led to speculation that molecules are destroyed in more energetic systems such that the outflowing gas is primarily ionised \citep{Fiore:2017}. Our results show that molecules can survive in outflows, even from the brightest quasars ($L_{\rm AGN}>10^{48}~{\rm ergs \, s^{-1}}$). These outflows have velocities comparable to the maximum outflow velocities of ionised gas for quasars with $L_{\rm AGN}\sim10^{47}$--$10^{48}~{\rm ergs \, s^{-1}}$ \citep{Fiore:2017}. This suggests that the molecular outflow velocities are maintained as the outflow expands out of the galaxy ($\sim10$~kpc) where the cold outflow may either stall or be destroyed, transitioning to a purely ionised phase, as predicted by radiation-hydrodynamical simulations \citep{Costa:2018a,Costa:2018b}. 

Future investigations will involve other gas tracers to determine the multi-phase properties of gas in these outflows. Such data could be used to estimate the outflow mass, which requires an understanding of the physical conditions of the outflowing gas. Additionally, the CO emission has been resolved for all of the lensed quasar systems in the data reported here and for most of the sample \citep{Stacey:2018b,Stacey:2021}: by taking advantage of lensing magnification, gravitational lens modelling (e.g. \citealt{Stacey:2021,Rizzo:2021}) of higher quality data could allow us to reconstruct the CO emission and diagnose the outflow kinetics. 

Although the difference in reddening and CO line width is striking, our work here involves only four quasars with $E_{B-V}>0.1$. Future work would particularly benefit from a systematic study of a larger sample to follow the evolution of quasars via the relationship between CO velocity, quasar obscuration, Eddington ratio and star formation rate. Such studies will be important to constrain the life-cycles of AGN and their role in shaping the evolution of massive galaxies.

\section*{Acknowledgements}

The authors thank the anonymous referee, Adam Schaefar, Simon White and Volker Springel for helpful discussions, and Michael Bremer for help with the NOEMA data reduction. HRS acknowledges funding from the European Research Council (ERC) under the European Union's Horizon 2020 research and innovation programme (LEDA: grant agreement No 758853). This research used SciPy, NumPy and Matplotlib packages for Python \citep{Virtanen:2020,Harris:2020,Hunter:2007}, and also the NASA/IPAC Extragalactic Database (NED), SIMBAD \citep{Wenger:2000} and VizieR (DOI : \url{10.26093/cds/vizier}) catalogue access tools. We made use of ALMA data with project codes 2018.1.01008.S, 2019.1.00948.S and 2019.1.00964.S. ALMA is a partnership of ESO (representing its member states), NSF (USA) and NINS (Japan), together with NRC (Canada), MOST and ASIAA (Taiwan), and KASI (Republic of Korea), in cooperation with the Republic of Chile. The Joint ALMA Observatory is operated by ESO, AUI/NRAO and NAOJ. We also used observations carried out under project number S17BV [001] with the IRAM NOEMA Interferometer. IRAM is supported by INSU/CNRS (France), MPG (Germany) and IGN (Spain).

\section*{Data Availability}

All observations reported in this work are publicly available in the ALMA archive (\url{https://almascience.eso.org/aq}) or IRAM Data Archive (\url{https://www.iram-institute.org/EN/content-page-386-7-386-0-0-0.html}) and all the analysis was performed with publicly available software. The data sets generated during the current study are available from the corresponding author upon reasonable request.



\bibliographystyle{mnras}
\bibliography{references} 



\appendix

\section{Supporting figures}

\begin{figure*}
\centering
\includegraphics[height=0.25\textwidth]{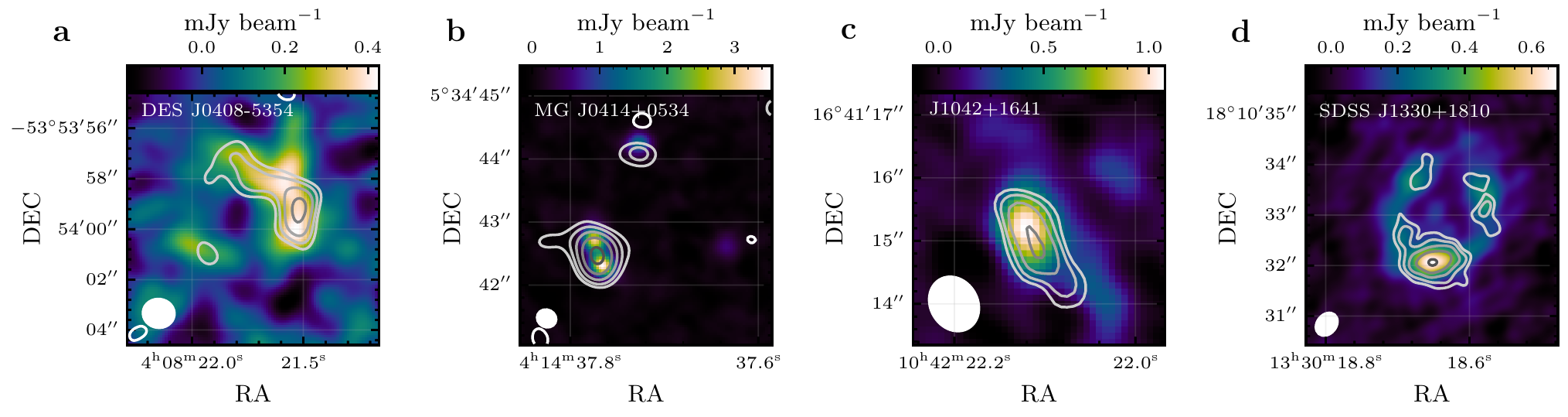}
\includegraphics[height=0.25\textwidth]{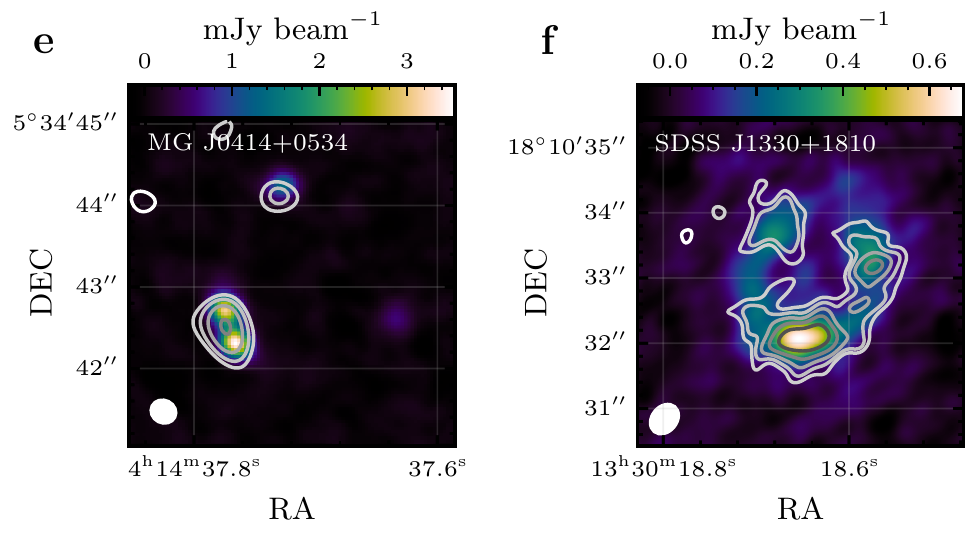}
    \caption{ALMA imaging for the new data presented in this work. {\bf a--d}: the sub-mm continuum emission with contours of the velocity-integrated CO line emission in signal-to-noise intervals of $-3,3,3\sqrt2,6,6\sqrt2$.. etc. The synthesised beam is shown by the ellipse in the lower-left corner. {\bf e, f}: the same for the [CI] line emission.}
\label{fig:images}
\end{figure*}

\begin{figure*}
    \centering
    \includegraphics[width=\textwidth]{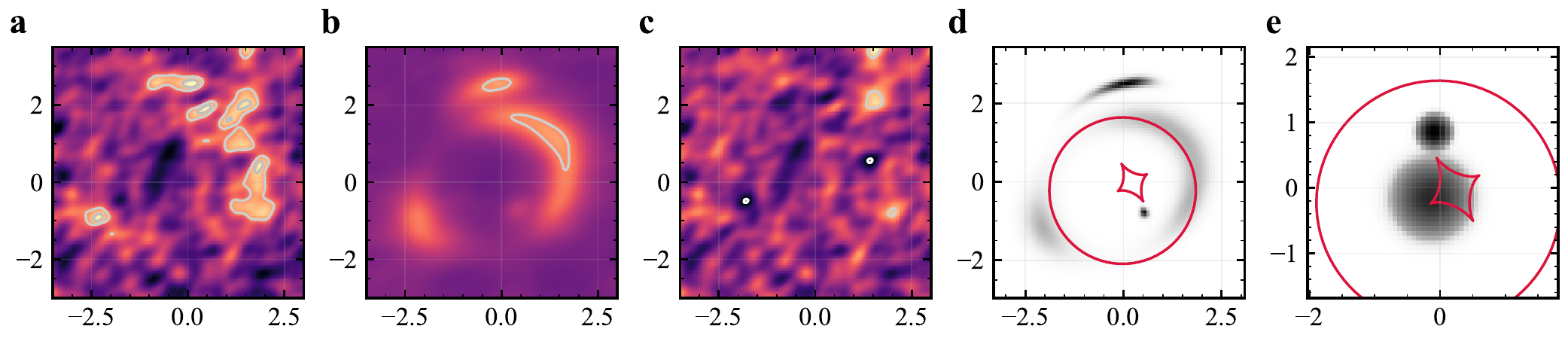}
    \caption{Lens modelling of DES\,J0408$-$5354 continuum. {\bf a--c}: The dirty image of the data, dirty image of the model (with the same colour-scale as the data) and residual image. The contours are in signal-to-noise intervals of $-3,3,3\sqrt2,6,6\sqrt2$.. etc. {\bf d}: The lens model (grey; log-scale) and lensing caustics (red). {\bf e} The source (grey; log-scale) and caustics (red). The scale is in arcsec relative to the phase centre of the observation. }
    \label{fig:0408_cont} \vspace{0.2cm}
    \includegraphics[width=\textwidth]{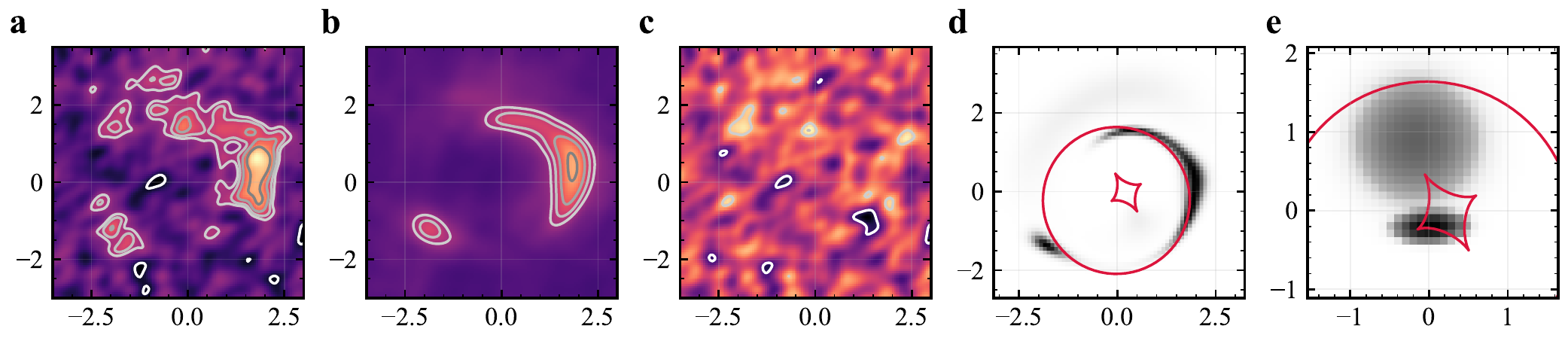}
    \caption{Lens modelling of DES\,J0408$-$5354 CO\,(7--6). Labels as in Fig.~\ref{fig:0408_cont}. }
    \label{fig:0408_line} \vspace{0.2cm}
    \includegraphics[width=\textwidth]{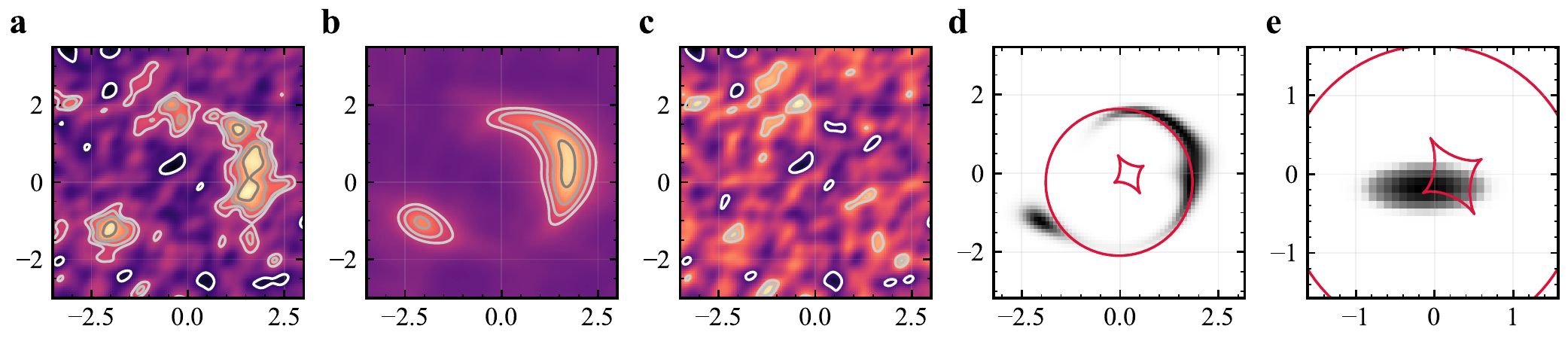}
    \caption{Lens modelling of DES\,J0408$-$5354 [CI]\,(2--1). Labels as in Fig.~\ref{fig:0408_cont}. }
    \label{fig:0408_line_CI}
\end{figure*}

\begin{figure*}
    \centering
    \includegraphics[width=\textwidth]{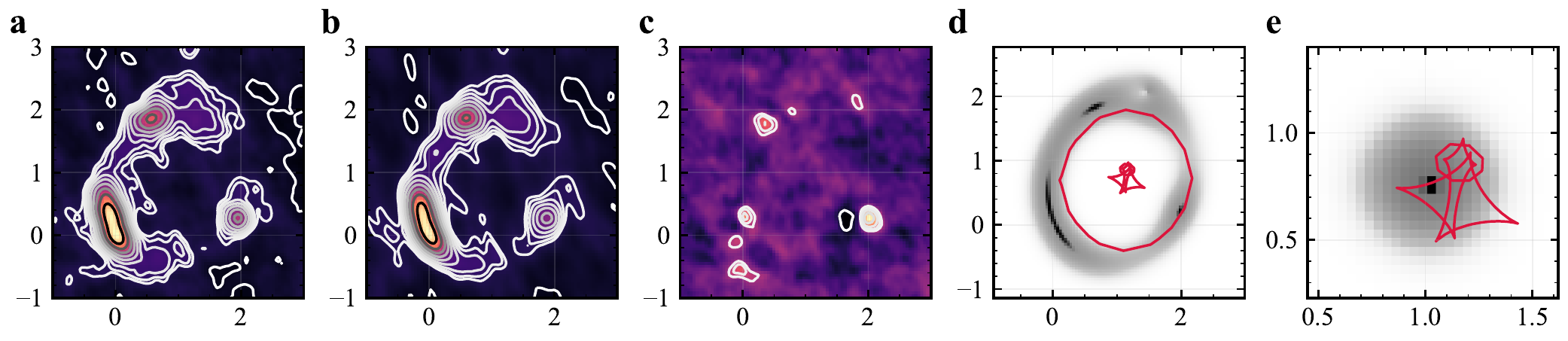}
    \caption{Lens modelling of MG\,J0414+0534 continuum. Labels as in Fig.~\ref{fig:0408_cont}. The synchrotron emission has been removed from the continuum (see text for explanation).}
    \label{fig:0414_cont} \vspace{0.2cm}
    \includegraphics[width=\textwidth]{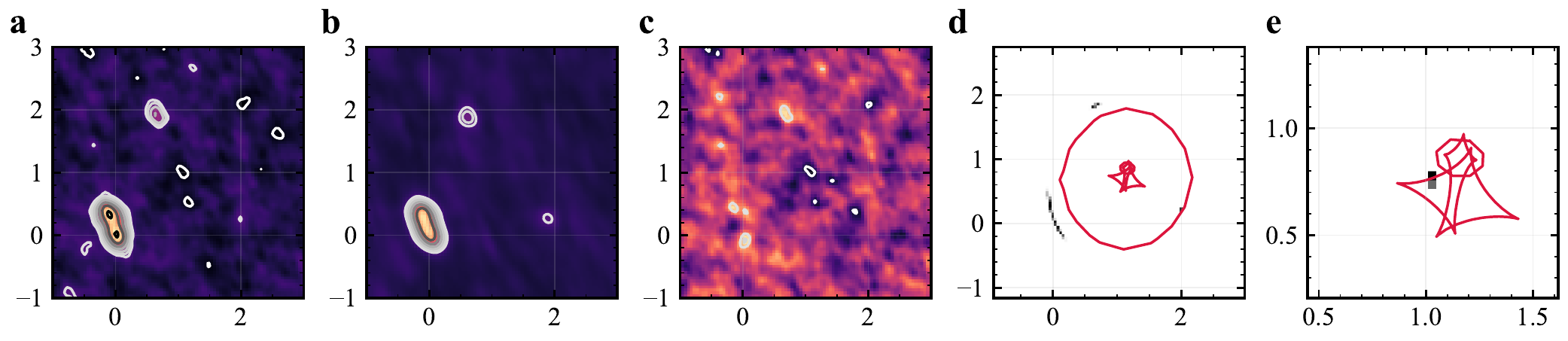}
    \caption{Lens modelling of MG\,J0414+0534 CO\,(11--10). Labels as in Fig.~\ref{fig:0408_line}.}
    \label{fig:0414_line}
\end{figure*}

\begin{figure*}
    \centering
    \includegraphics[width=\textwidth]{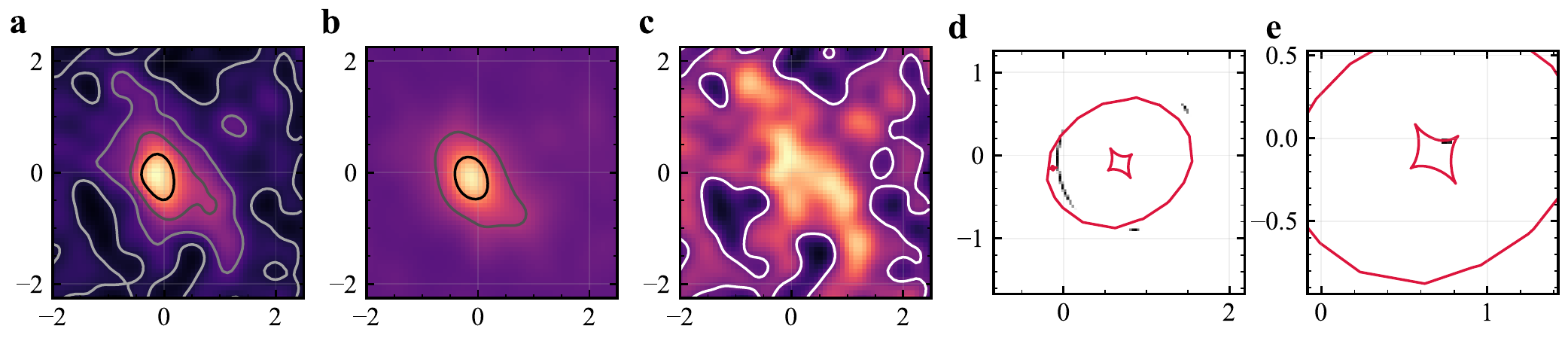}
    \caption{Lens modelling of J1042+1641 continuum. Labels as in Fig.~\ref{fig:0408_cont}.}
    \label{fig:1042_cont}
    \includegraphics[width=\textwidth]{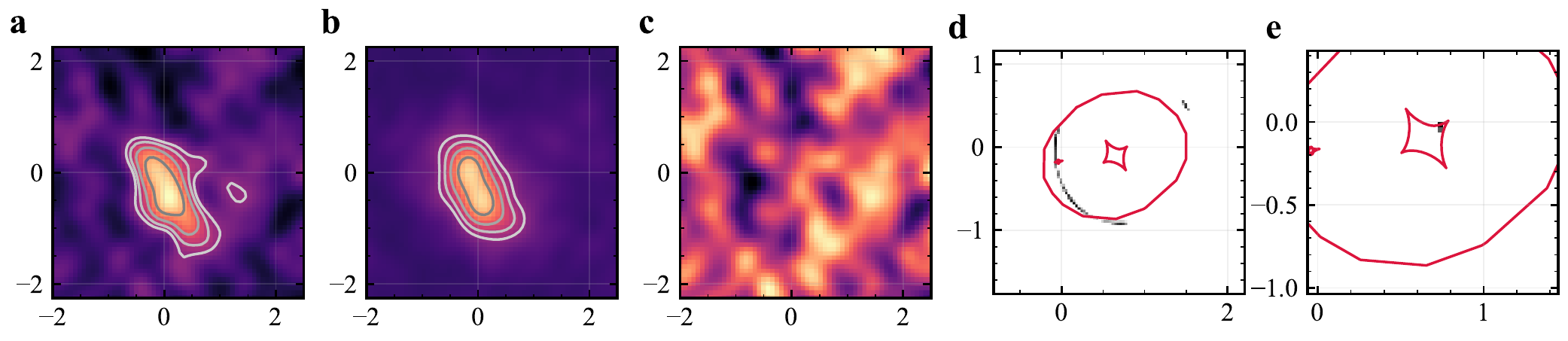}
    \caption{Lens modelling of J1042+1641 CO\,(10--9). Labels as in Fig.~\ref{fig:0408_cont}.}
    \label{fig:1042_line}
\end{figure*}

\begin{figure*}
    \centering
    \includegraphics[width=\textwidth]{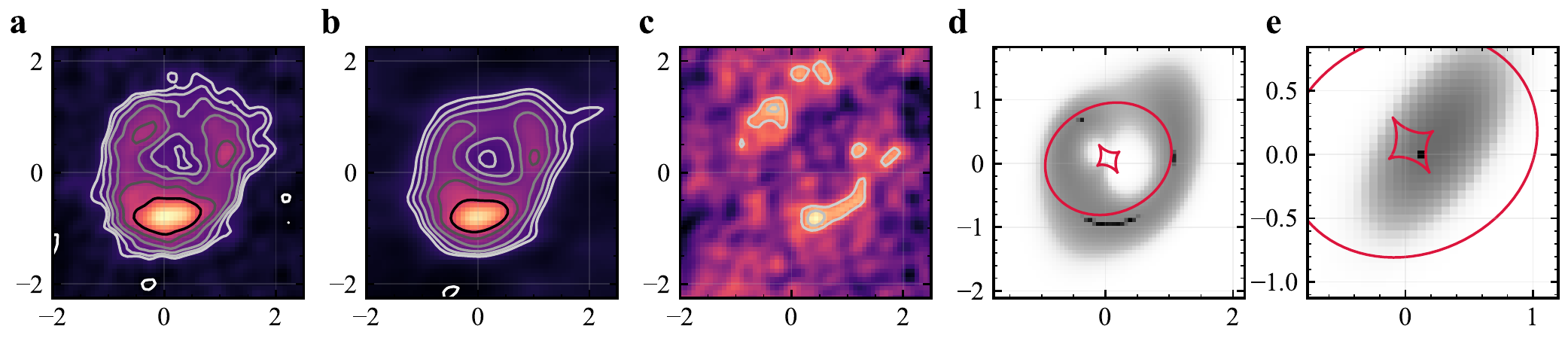}
    \caption{Lens modelling of SDSS\,J1330+1810 continuum. Labels as in Fig.~\ref{fig:0408_cont}.}
    \label{fig:1330_cont}
    \includegraphics[width=\textwidth]{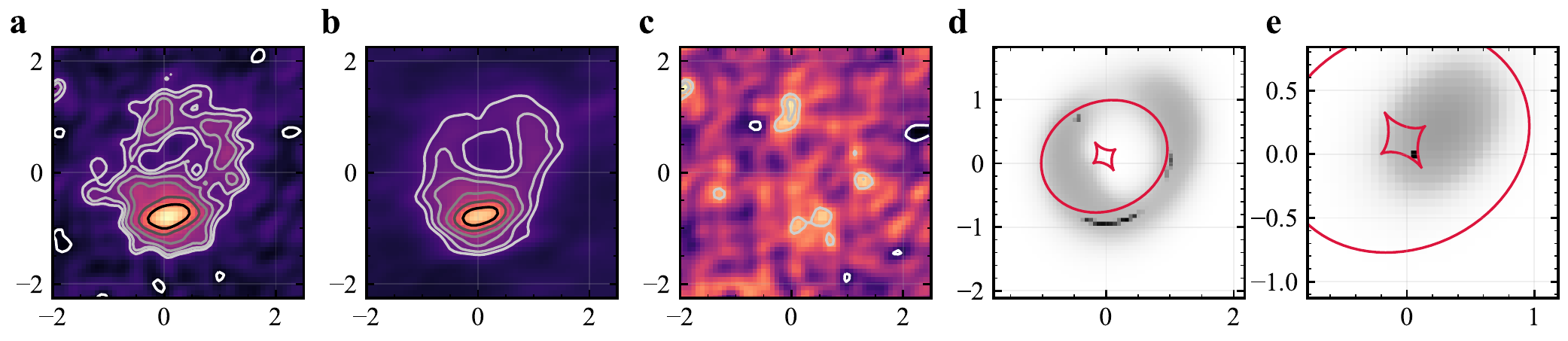}
    \caption{Lens modelling of SDSS\,J1330+1810 CO\,(7--6). Labels as in Fig.~\ref{fig:0408_cont}.}
    \label{fig:1330_CO}
    \includegraphics[width=\textwidth]{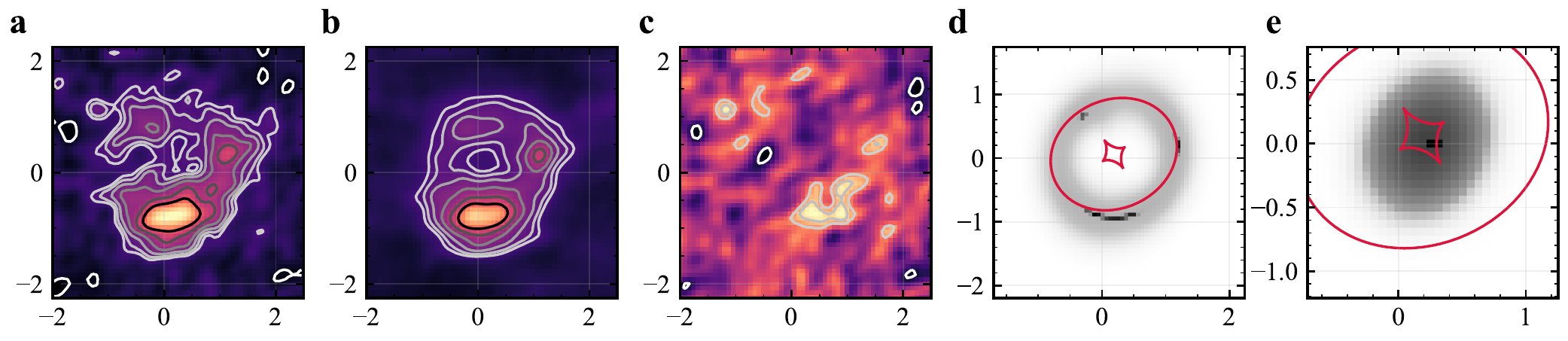}
    \caption{Lens modelling of SDSS\,J1330+1810 CI\,(2--1). Labels as in Fig.~\ref{fig:0408_cont}.}
    \label{fig:1330_CI}
\end{figure*}

\begin{figure*} 
\begin{tikzpicture}
    \centering
    
    \draw (0, 0) node[inner sep=0] 
    {\includegraphics[width=0.245\textwidth]{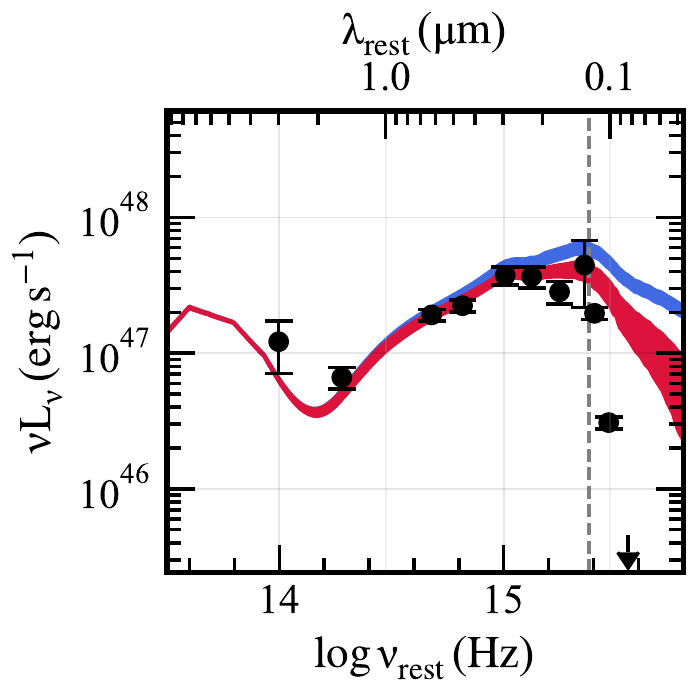}};
    \draw (0.3, 1.1) node {\small SDSS\,J0100$+$2802};

    \draw (4.3, 0) node[inner sep=0] {\includegraphics[width=0.245\textwidth]{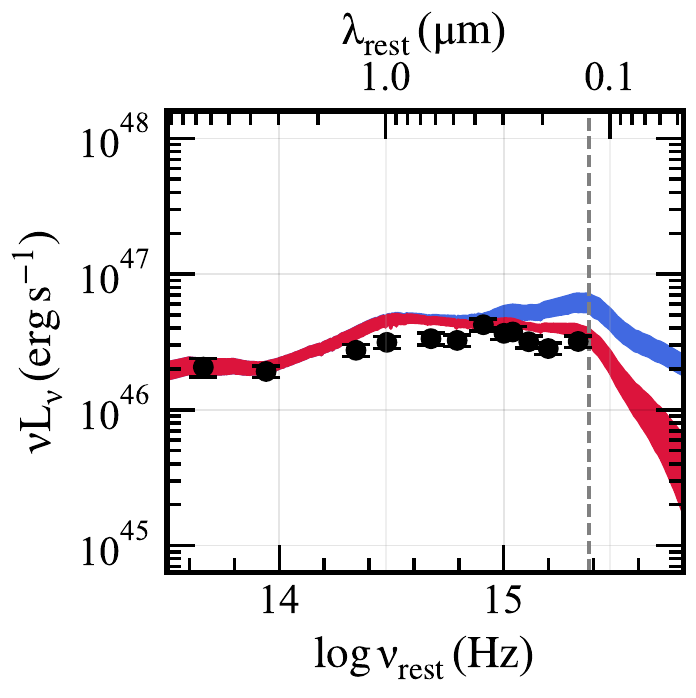}};
    \draw (4.6, 1.1) node {\small DES\,J0408$-$5354};

    \draw (8.6, 0) node[inner sep=0] {\includegraphics[width=0.245\textwidth]{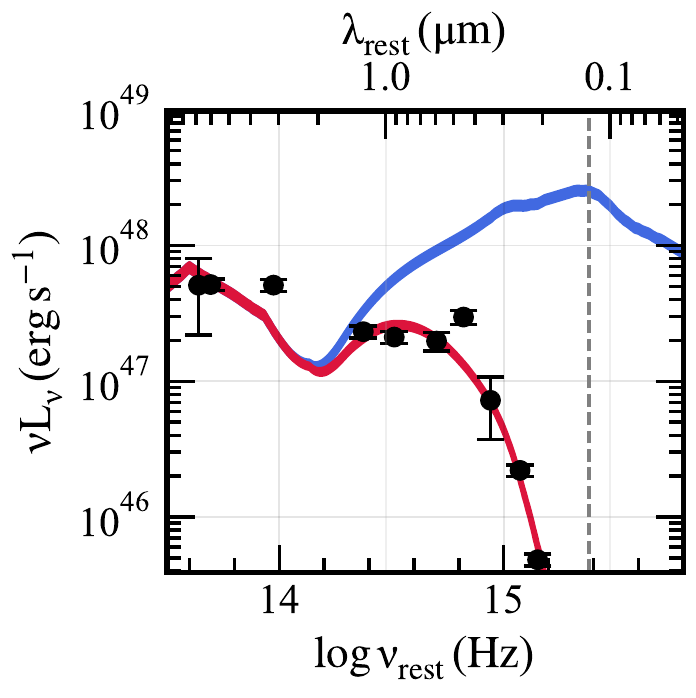}};
    \draw (8.8, 1.1) node {\small MG\,J0414$+$0534};

    \draw (12.9, 0) node[inner sep=0] {\includegraphics[width=0.245\textwidth]{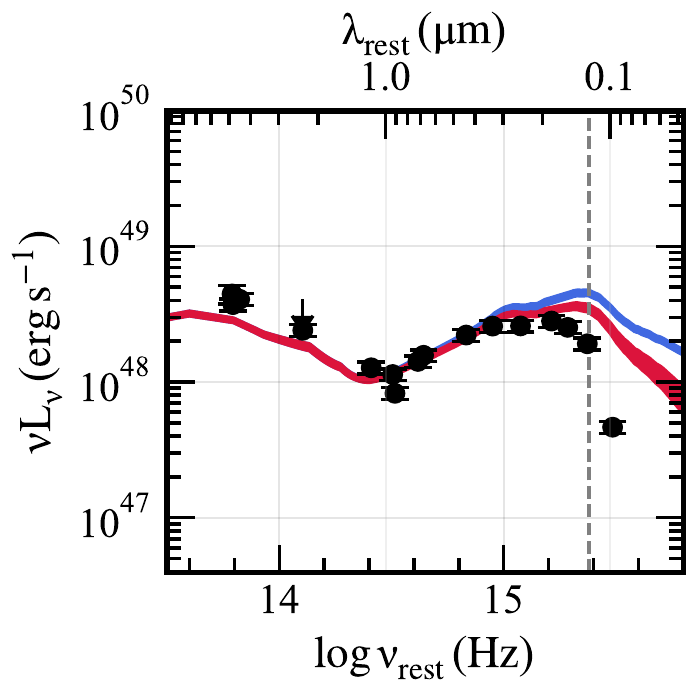}};
    \draw (13.2, 1.1) node {\small APM\,08279$+$5255};

    \draw (0, -4.5) node[inner sep=0] {\includegraphics[width=0.245\textwidth]{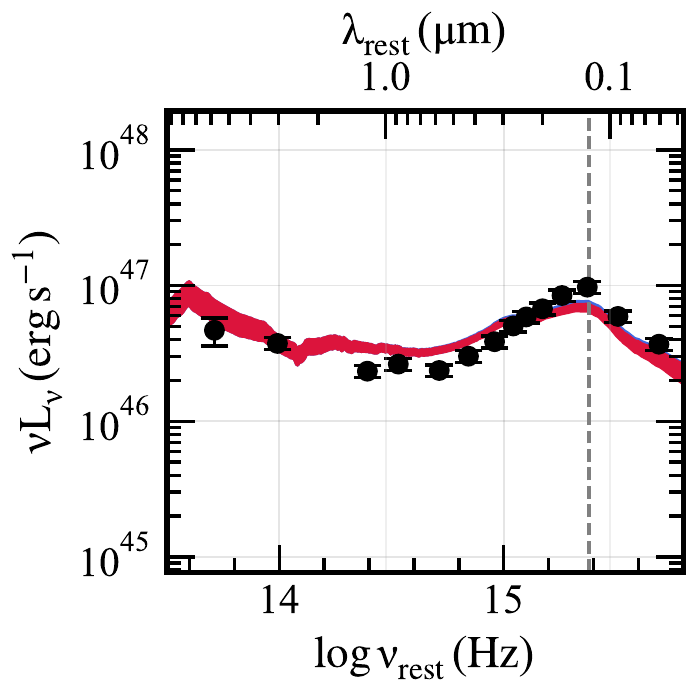}};
    \draw (0.2, -3.4) node {\small RX\,J0911$+$0551};  

     \draw (4.3, -4.5) node[inner sep=0] {\includegraphics[width=0.245\textwidth]{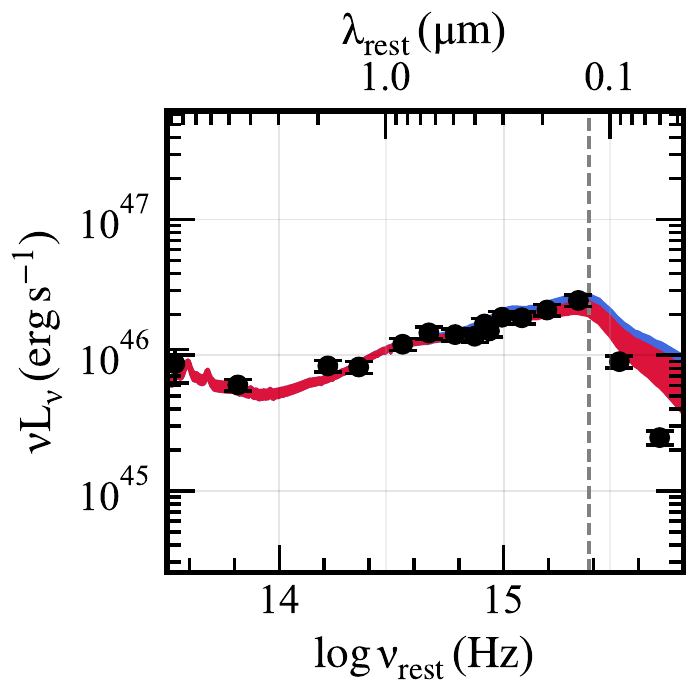}};
    \draw (4.6, -3.4) node {\small SDSS\,J0924$+$0219};  

     \draw (8.6, -4.5) node[inner sep=0] {\includegraphics[width=0.245\textwidth]{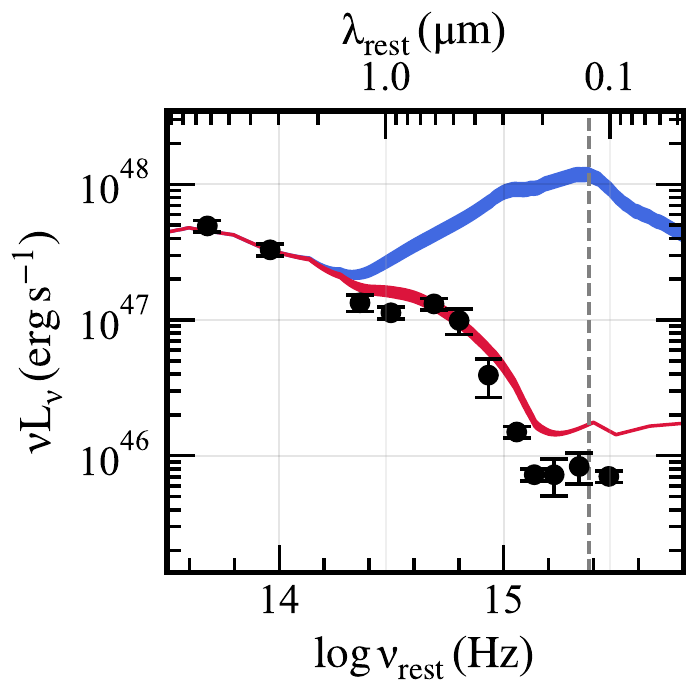}};
    \draw (8.6, -3.4) node {\small J1042$+$1641};  

     \draw (12.9, -4.5) node[inner sep=0] {\includegraphics[width=0.245\textwidth]{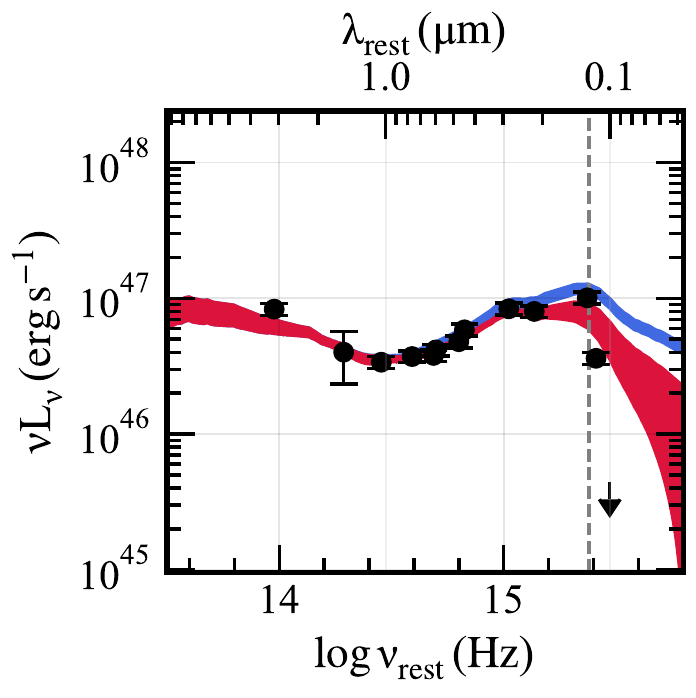}};
    \draw (13.2, -3.4) node {\small SDSS\,J1148$+$5251};  
    
     \draw (0.0, -9.0) node[inner sep=0] {\includegraphics[width=0.245\textwidth]{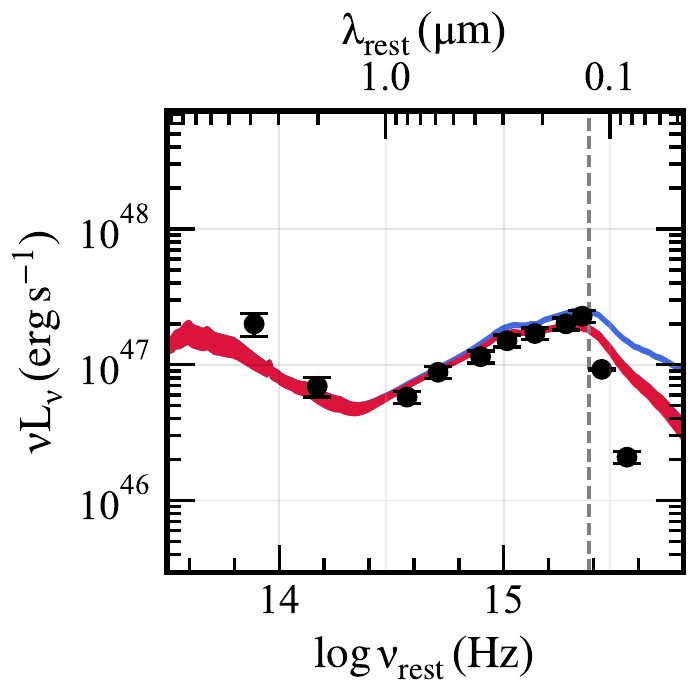}};
    \draw (0.15, -7.9) node {\small BRI\,1202$-$0725};  

     \draw (4.3, -9.0) node[inner sep=0] {\includegraphics[width=0.245\textwidth]{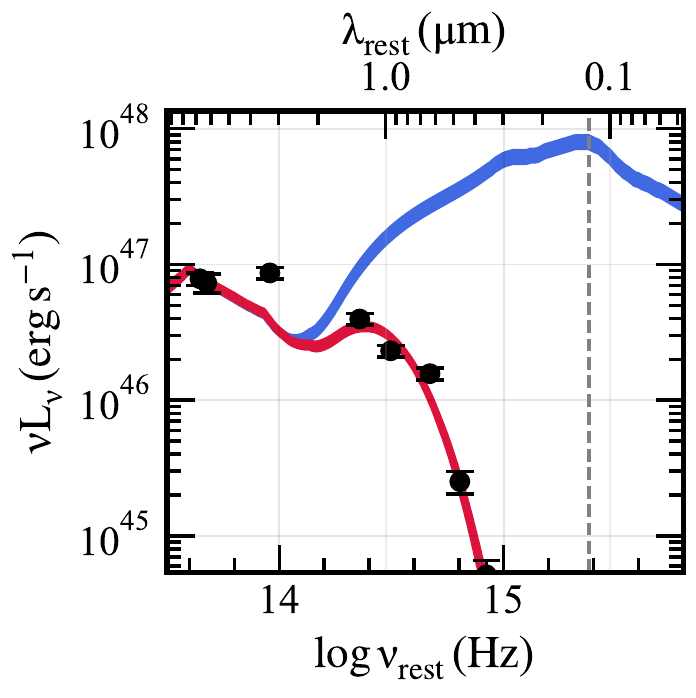}};
    \draw (4.6, -7.9) node {\small ULAS\,J1234$+$0907};  

     \draw (8.6, -9.0) node[inner sep=0] {\includegraphics[width=0.245\textwidth]{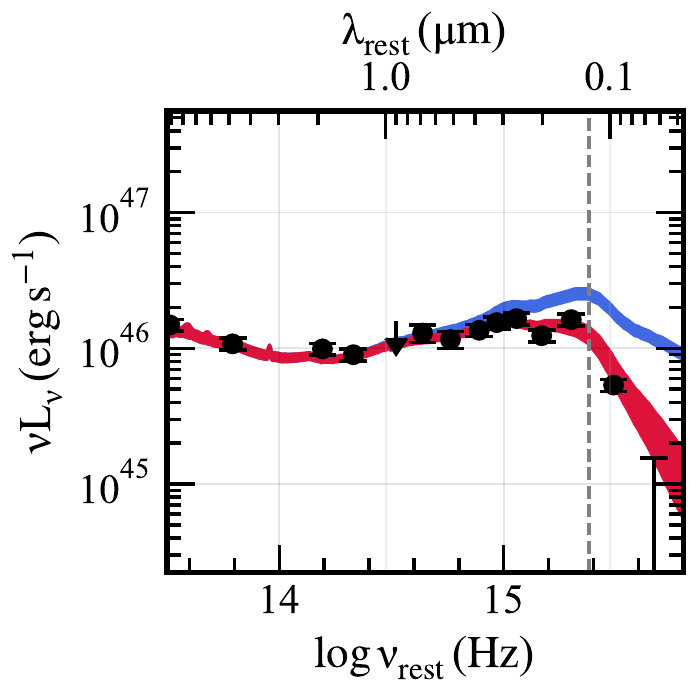}};
    \draw (8.9, -7.9) node {\small SDSS\,J1330$+$1810};  

     \draw (12.9, -9.0) node[inner sep=0] {\includegraphics[width=0.245\textwidth]{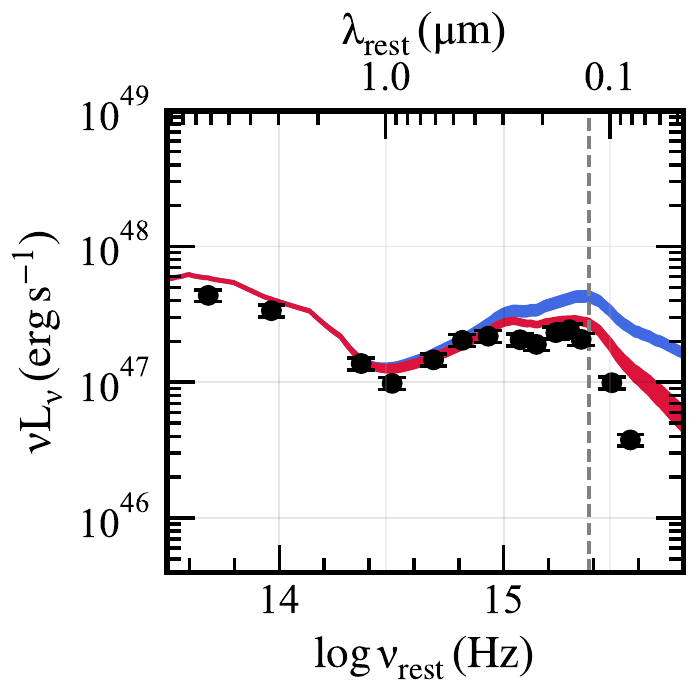}};
    \draw (12.9, -7.9) node {\small H1413$+$117};  

     \draw (4.3, -13.5) node[inner sep=0] {\includegraphics[width=0.245\textwidth]{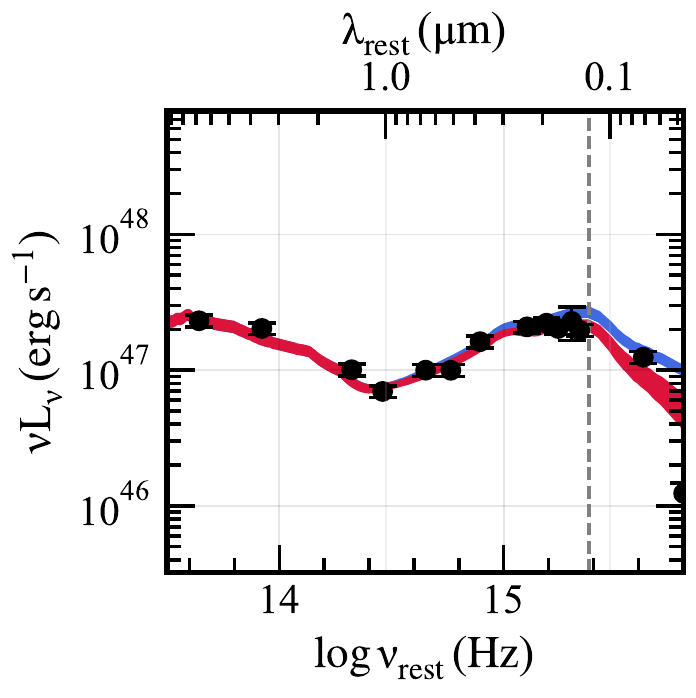}};
    \draw (4.5, -12.4) node {\small WFI\,J2026$-$4536};  

     \draw (0.0, -13.5) node[inner sep=0] {\includegraphics[width=0.245\textwidth]{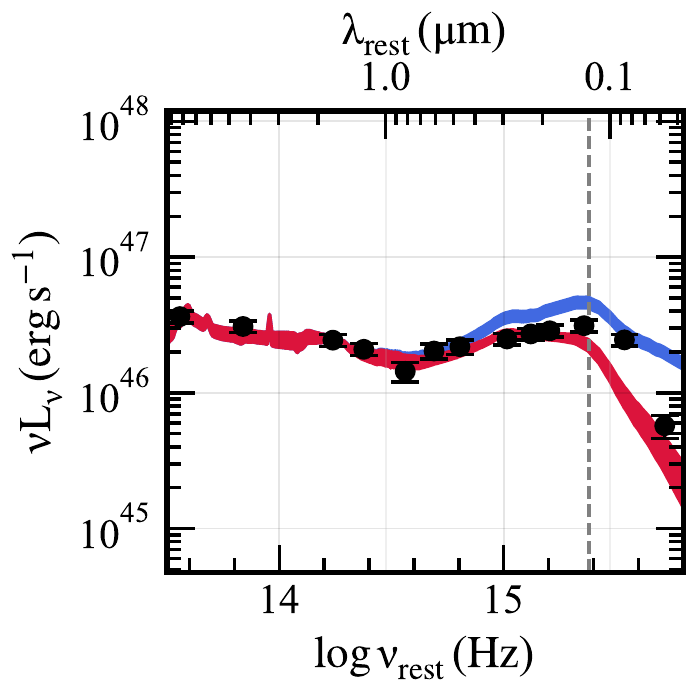}};
    \draw (0.2, -12.4) node {\small WFI\,J2033$-$4723};  

     \draw (4.3, -13.5) node[inner sep=0] {\includegraphics[width=0.245\textwidth]{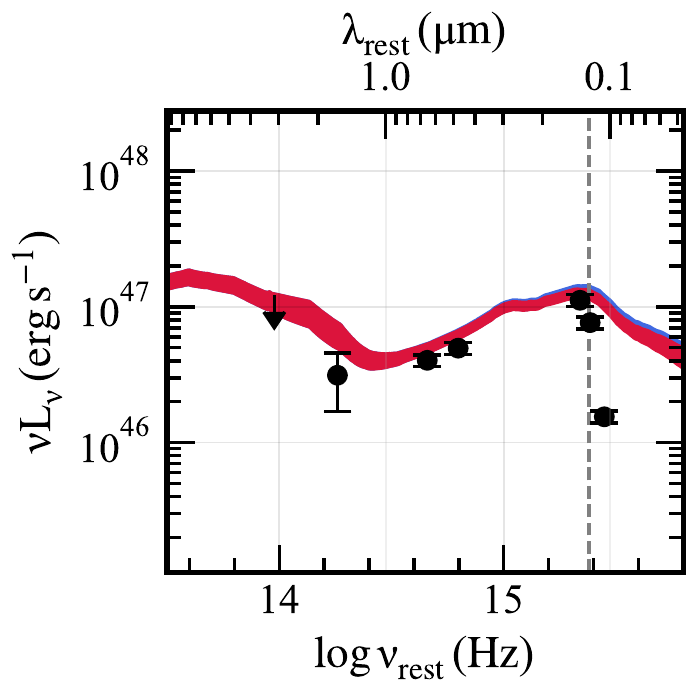}};
    \draw (4.6, -12.4) node {\small SDSS\,J2310$+$1814};  
    
     \draw (8.6, -13.5) node[inner sep=0] {\includegraphics[width=0.245\textwidth]{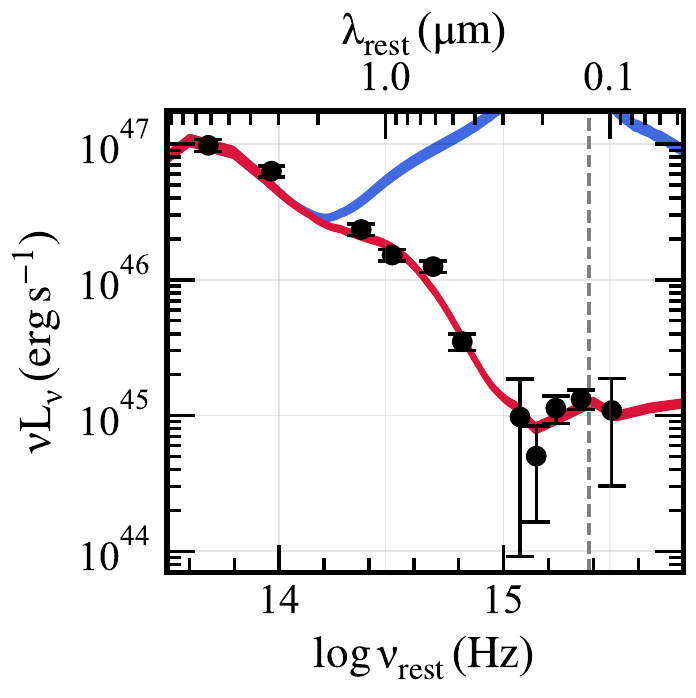}};
    \draw (8.9, -12.4) node {\small ULAS\,J2315$+$0143};  

\end{tikzpicture}
    \caption{ SED fits for the quasar sample. We use AGNFitter \citep{Calistro:2016} to fit broad-band SEDs from the ultra-violet to mid-infrared. The red curve is a solid fill between 16th to 86th percentiles of the models from the MCMC analysis. The blue curve shows the quasar models without dust attenuation. The grey dashed line is the rest-frame frequency of Lyman-$\upalpha$; photometric measurements at or above this frequency are ignored because they are affected by absorption by the intergalactic medium. }
    \label{fig:SEDs}
\end{figure*}

\section{Supporting tables}

\begin{table*}
    \caption{ Summary of lens modelling in this work, using {\sc visilens} \citep{Hezaveh:2013,Spilker:2016}. We give the marginalised posteriors from the MCMC of the major axis ($a$), axis ratio ($q$), magnification ($\mu$) for each Gaussian source component. The lens model parameters are kept fixed except the position of the lens, to account for any astrometric offsets. $^\ddagger$ After removal of synchrotron emission, see text for details. }
    \centering
    \setlength{\tabcolsep}{3pt}
    \adjustbox{width=\textwidth}{
    \renewcommand*{\arraystretch}{1.1}
    \begin{tabular}{l c c c c c c c c c l} \hline
        Name & Data & $a_1$ & $a_1$ & $q_1$ & $\mu_1$ & $a_2$ & $a_2$ & $q_2$ & $\mu_2$ & Lens model \\ 
          &   & (arcsec) & (kpc) & & & (arcsec) & (kpc) & & & reference  \\ \hline \noalign{\vskip 0.1cm} 
        MG\,J0414+0534 & continuum$^\ddagger$ & $0.019\pm0.001$ & $0.16\pm0.01$ & $\equiv1$ & $29.6\pm0.3$ & $0.189\pm0.005$ & $1.56\pm0.04$ & $\equiv1$ & $15.4\pm0.2$ & \cite{Stacey:2020} \\
                    & CO\,(11--10) & $0.009\pm0.002$ & $0.07\pm0.02$ & $\equiv1$ & $42\pm5$ & - & - & - & - & \\ 
        DES\,J0408$-$5354 & continuum & $0.30\pm0.07$ & $2.5\pm0.6$ & $\equiv 1$ & $10\pm1$ & $0.14\pm0.09$ & $1.2\pm0.8$ & $\equiv 1$ & $6\pm4$ & \cite{Agnello:2017} \\
                    & CO\,(7--6) & $0.21\pm0.02$ & $1.8\pm0.2$ & $0.4\pm0.1$ & $18\pm1$ & $0.46\pm0.04$ & $3.8\pm0.3$ & $\equiv1$ & $5.1\pm0.6$ &  \\
                    & [CI]\,(2--1) & $0.31\pm0.02$ & $2.6\pm0.2$ & $0.4\pm0.1$ & $14\pm1$ & - & - & - & - & \\
        J1042$+$1641 & continuum & $0.008\pm0.002$ & $0.07\pm0.02$ & $\equiv1$ & $49\pm7$ & - & - & - & - & \cite{Glikman:2018} \\
                    & CO\,(10--9) & $0.008\pm0.001$ & $0.07\pm0.01$ & $\equiv1$ & $43\pm1$ & - & - & - & - & \\ 
        SDSS\,J1330+1810 & continuum & $0.014\pm0.003$ & $0.12\pm0.02$ & $\equiv1$ & $26\pm1$ & $0.46\pm0.02$ & $4.0\pm0.2$ & $0.48\pm0.03$ & $6.8\pm0.3$ & \cite{Shajib:2019} \\
                    & CO\,(7--6) & $0.014\pm0.001$ & $0.12\pm0.01$ & $\equiv1$ & $24\pm1$ & $0.44\pm0.05$ & $3.8\pm0.4$ & $0.7\pm0.1$ & $5.6\pm0.3$ & \\
                    & [CI]\,(2--1) & $0.014\pm0.001$ & $0.12\pm0.01$ & $\equiv1$ & $23\pm1$ & $0.34\pm0.05$ & $2.9\pm0.4$ &  $0.8\pm0.1$ & $8.4\pm0.8$ & \\ \noalign{\vskip 0.1cm} \hline
    \end{tabular}}
    \label{table:lensmodels}
\end{table*}

\begin{landscape}
\begin{table}
    \caption{Summary of the objects in the sample. We give the number of lensed quasar images ($N_{\rm im}=1$ if not strongly lensed), source redshift ($z_s$), CO line transition (or other line), line FWHM from a single Gaussian fit to the line profile, SFR (lensing-corrected), black hole mass ($M_{\rm BH}$), extinction ($E_{B-V}$), integrated luminosity of the accretion disc template ($\log\,L_{\rm 0.05\--1\upmu m}$, {\it not} lensing-corrected) and assumed value of the quasar magnification from the literature ($\mu_{\rm qso}$). Where no uncertainty was given for the black hole mass, we assume an uncertainty of 0.25 dex which is the typical scatter found for black hole scaling relations. If no uncertainty is given for the star formation rate, we assume it is only accurate to within a factor of 2. References given are for the line FWHM, black hole mass and quasar magnification, and photometry used for the optical--infrared SED fitting (if not from all-sky catalogues). $^{\ddagger}$A fit to the CO FWHM was not reported for this line, but reported to be consistent with the [CII] FWHM.} 
    \centering
    \setlength{\tabcolsep}{5pt}
    \adjustbox{width=1.33\textwidth}{
    \renewcommand*{\arraystretch}{1.1}
    \begin{tabular}{l c c c c c c c c c l} \hline \noalign{\vskip 0.1cm} 
        Name & $N_{\rm im}$ & $z_{\rm s}$ & Line & $\Theta_{\rm line}$ & SFR & $\log M_{\rm BH}$ & $E_{B-V}$ & $\log L_{\rm 0.05\--1\upmu m}$ & $\mu_{\rm qso}$ & References \\
        & & & & (\kms) & (\sfr) & (\msol) & & (ergs\,s$^{-1}$) &  & \\ \noalign{\vskip 0.1cm} \hline \noalign{\vskip 0.1cm} 
        DES\,J0408$-$5354 & 4 & 2.375 & CO\,(7--6) & $260\pm100$ & $240^{+240}_{-120}$ & $8.41\pm0.27$ & $0.038_{-0.016}^{+0.015}$ & $47.04_{-0.08}^{+0.07}$ & 22 & \cite{Shajib:2020}, this work \\
         & & & [CI]\,(2--1) & $140\pm40$ & & & & & & This work \\
        MG\,J0414$+$0534 & 4 & 2.64 & CO\,(11--10) & $1080\pm20$ & $800\pm70$ & 9.26 & $0.738_{-0.014}^{+0.017}$ & $48.66_{-0.03}^{+0.04}$ & 56.4 & \cite{Blackburne:2011,Stacey:2018a,Stacey:2018b} \\ 
         & & & CO\,(7--6) & $1390\pm240$ & & & & & &  This work \\
         & & & [CI]\,(2--1) & $500\pm120$ & & & & & &  This work \\
         & & & CO\,(3--2) & $470\pm40$ & & & & & &  This work \\
        APM\,08279$+$5255 & 4 & 3.91 & CO\,(9--8) & $460\pm20$ & $10000\pm200$ & 10.0 & $0.018_{-0.006}^{+0.006}$ & $48.91_{-0.02}^{+0.01}$ & 4 &  \cite{Weiss:2007,Stacey:2018a} \\
                        & & & [CI]\,(1--0) & $390\pm70$ & & & & & & \cite{Wagg:2006} \\
        RX\,J0911$+$0551 & 4 & 2.79 & CO\,(11--10) & $100\pm30$ & $750\pm50$ & 8.9 & $0.002_{-0.002}^{+0.005}$ & $47.10_{-0.02}^{+0.02}$ & 17.5 & \cite{Blackburne:2011,Stacey:2021,Tuan-Anh:2017} \\
                        & & & CO\,(3--2) & $360\pm60$ & & & & & & \cite{Tuan-Anh:2017} \\
        SDSS\,J0924$+$0219 & 4 & 1.525 & CO\,(8--7) & $180\pm10$ & $60\pm20$ & $7.93\pm0.35$ & $0.008_{-0.006}^{+0.012}$ & $46.65_{-0.02}^{+0.04}$ & 26.2 &  \cite{Sluse:2012} \\
        J1042$+$1641 & 4 & 2.25 & CO\,(10--9) & $700\pm120$ & $140^{+140}_{-70}$ & $9.31\pm0.25$ & $0.65_{-0.02}^{+0.02}$ & $48.34_{-0.05}^{+0.02}$ & 117 &  \cite{Glikman:2018}, this work \\
        SDSS\,J1330$+$1810 & 4 & 1.39 & CO\,(7--6) & $190\pm20$ & $150^{+160}_{-80}$ & - & $0.05_{-0.02}^{+0.01}$ & $46.66_{-0.04}^{+0.04}$ & 68.2 & \cite{Blackburne:2011,Stacey:2018a}, this work \\
         & & & [CI]\,(2--1) & $158\pm5$ & & & & &  & This work \\
        H1413$+$117 & 4 & 2.56 & CO\,(9--8) & $370\pm10$ & $1600\pm60$ & $9.12\pm0.01$ & $0.06_{-0.02}^{+0.06}$ & $47.93_{-0.04}^{+0.04}$ & 10.3 & \cite{Stacey:2021,Sluse:2012} \\
                        & & & CO\,(3--2) & $360\pm20$ & & & & & & \cite{Barvainis:1997} \\
        WFI\,J2026$-$4536 & 4 & 2.23 & CO\,(10--9) & $252\pm7$ & $1010\pm60$ & $9.18\pm0.36$ & $0.02_{-0.01}^{+0.01}$ & $47.68_{-0.02}^{+0.04}$ & 33.0 & \cite{Stacey:2021,Cornachione:2020,Bate:2018} \\
        WFI\,J2033$-$4723 & 4 & 1.66 & CO\,(8--7) & $80\pm10$ & $100\pm30$ & $8.63\pm0.35$ & $0.03_{-0.03}^{+0.05}$ & $47.08_{-0.04}^{+0.06}$ & 19.7 & \cite{Stacey:2021,Sluse:2012,Morgan:2004} \\
        SDSS\,J0100$+$2802 & 1 & 6.33 & CO\,(10--9) & 380$^{\ddagger}$ & $2800\pm1400$ & $10.09\pm0.07$ & $0.02_{-0.01}^{+0.02}$ & $47.98_{-0.04}^{+0.06}$ & - & \cite{Wu:2015,Wang:2019} \\
        & & & [CII] & $380\pm20$ & & & & & & \cite{Wang:2019} \\
        SDSS\,J1148$+$5251 & 1 & 6.42 & CO\,(7--6) & $300\pm40$ & $3600\pm1200$ & $9.43\pm0.07$ & $0.02_{-0.01}^{+0.02}$ & $47.30_{-0.03}^{+0.04}$ & - & \cite{Carniani:2019,Leipski:2013,Gallerani:2017} \\
         & & & [CII] & $287\pm28$ & & & & & & \cite{Walter:2009}\\
        BRI1202$-$0725 & 1 & 4.70 & CO\,(7--6) & $316\pm40$ & $3200\pm1600$ & $9.1\pm0.3$ & $0.020^{+0.005}_{-0.005}$ & $47.66^{+0.01}_{-0.01}$ & - & \cite{Salome:2012,Carniani:2013} \\
                &   &      & [CII] & $275\pm15$ & & & & & & \cite{Carilli:2013} \\
        ULAS\,J1234$+$0907 & 1 & 2.50 & CO\,(7--6) & $900\pm80$ & $2500\pm900$ & 10.4 & $1.77 _{-0.07}^{+0.07}$ & $48.16_{-0.04}^{+0.05}$ & - & \cite{Banerji:2017,Banerji:2018}\\
         & & & CO\,(3-2) & $870\pm60$ & & & & & & \cite{Banerji:2021} \\
        SDSS\,J2310$+$1814 & 1 & 6.00 & CO\,(9--8) & $380\pm20$ & $4300\pm700$ & 9.6 & $0.004_{-0.003}^{+0.007}$ & $47.37_{-0.03}^{+0.04}$ & - & \cite{Carniani:2019,Li:2020,Feruglio:2018} \\
         & & & [CII] & $390\pm20$ & & & & & & \cite{Li:2020} \\ 
         ULAS2315$+$0143 & 1 & 2.57 & CO\,(7--6) & $580\pm70$ & $3000\pm600$ & 10.1 & $1.27^{+0.07}_{-0.08}$ & $47.66^{+0.05}_{-0.05}$ & - & \cite{Banerji:2017,Banerji:2018} \\ 
                & & & CO\,(3--2) & $300\pm60$ & & & & & & \cite{Banerji:2021} \\ \hline
    \end{tabular} }
\label{table:sources}
\end{table}
\end{landscape}


\bsp	
\label{lastpage}
\end{document}